\documentclass[fleqn,usenatbib]{mnras}

\usepackage{newtxtext,newtxmath}
\usepackage{adjustbox}

\usepackage[T1]{fontenc}

\DeclareRobustCommand{\VAN}[3]{#2}
\let\VANthebibliography\thebibliography
\def\thebibliography{\DeclareRobustCommand{\VAN}[3]{##3}\VANthebibliography}

\usepackage{graphicx}	
\usepackage{amsmath}	

\title[New ATLAS mCP stars]{New magnetic chemically peculiar stars and candidates in the ATLAS First Catalog of Variable Stars}
\author[K.~Bernhard et al.]{
Klaus Bernhard,$^{1,2}$\thanks{E-mail: klaus.bernhard@liwest.at (KB)}
Stefan H{\"u}mmerich,$^{1,2}$
Ernst Paunzen$^{3}$ and
Johana Sup{\'i}kov{\'a}$^{4}$
\\
$^{1}$Bundesdeutsche Arbeitsgemeinschaft f{\"u}r Ver{\"a}nderliche Sterne e.V. (BAV), Munsterdamm 90, D-12169 Berlin, Germany\\
$^{2}$American Association of Variable Star Observers (AAVSO), 49 Bay State Rd, Cambridge, MA 02138, USA\\
$^{3}$Department of Theoretical Physics and Astrophysics, Masaryk University, Kotl\'a\v{r}sk\'a 2, 611\,37 Brno, Czech Republic\\
$^{4}$Faculty of Informatics, Masaryk University, Botanick\'a 68A, 602 00 Brno, Czech Republic}

\date{Accepted XXX. Received YYY; in original form ZZZ}

\pubyear{2021}

\begin{document}
\label{firstpage}
\pagerange{\pageref{firstpage}--\pageref{lastpage}}
\maketitle

\begin{abstract} 
The number of known variable stars has increased by several magnitudes over the last decade, and automated classification routines are becoming increasingly important to cope with this development. Here we show that the 'upside-down CBH variables', which were proposed as a potentially new class of variable stars by \citet{heinze18} in the ATLAS First Catalogue of Variable Stars, are, at least to a high percentage, made up of $\alpha^{2}$ Canum Venaticorum (ACV) variables -- that is, photometrically variable magnetic chemically peculiar (CP2/He-peculiar) stars -- with distinct double-wave light curves. Using suitable selection criteria, we identified 264 candidate ACV variables in the ATLAS variable star catalogue. 62 of these objects were spectroscopically confirmed with spectra from the Large Sky Area Multi-Object Fiber Spectroscopic Telescope (all new discoveries except for nine stars) and classified on the MK system. The other 202 stars are here presented as ACV star candidates that require spectroscopic confirmation. The vast majority of our sample of stars are main-sequence objects. Derived masses range from 1.4\,\(\textup{M}_\odot\) to 5\,\(\textup{M}_\odot\), with half our sample stars being situated in the range from 2 \,\(\textup{M}_\odot\) to 2.4 \,\(\textup{M}_\odot\), in good agreement with the spectral classifications. Most stars belong to the thin or thick disk; four objects, however, classify as members of the halo population. With a peak magnitude distribution at around 14th magnitude, the here presented stars are situated at the faint end of the known Galactic mCP star population. Our study highlights the need to consider rare variability classes, like ACV variables, in automated classification routines.
\end{abstract}




\section{Introduction} \label{introduction}

Over the last decades, rapid advances in technology and the resulting increasing number of large-scale digital sky surveys have helped to propel astronomy into a new data-rich era \citep{djorgovski13}. This development necessitated the invention of novel ways to deal with the amount of incoming data produced by the petabyte-scale sky surveys, and algorithmic advances in the automated handling of massive and complex data sets have become of great importance.

The field of variable star research is no exception and has greatly benefited from the increasing amount of available photometric time-series data, which are often made available in real time. The number of known variable stars has increased by several magnitudes over the last decade alone, with recent variable star catalogues containing hundreds of thousands of entries (e.g. \citealt{heinze18,ZTFvars}) and the International Variable Star Index (VSX; \citealt{VSX}) of the American Association of Variable Star Observers (AAVSO), the most coherent and up-to-date database of variable stars, cataloging more than 2,100,000 objects at the time of this writing (June 2021). Automated classification routines are of essential importance to sort the vast number of newly discovered variable stars into astrophysically meaningful classes. However, with the development of new and improved algorithms and classification networks, new (meta)classes are constantly developed, and it is not easy to keep track with this steadily growing body of nomenclature.

In agreement with the general trend, an increasingly large number of photometrically variable chemically peculiar (CP) stars have been identified during the past decades \citep[e.g.][]{paunzen98,wraight12,bernhard15b,bernhard15a,huemmerich16,bowman18,huemmerich18,david_uraz19,sikora19}. CP stars are upper main-sequence objects, which are generally found between spectral types early B to early F and characterised by peculiar surface abundances. The observed peculiarities are thought to have their origin in the interplay between radiative levitation and gravitational settling (atomic diffusion) in the calm outer layers of slowly rotating stars (e.g. \citealt{michaud70,richer00}). Several types of CP stars exist, such as the metallic-line or Am/CP1 stars, the magnetic Ap/CP2 stars, the HgMn/CP3 stars, and the He-peculiar stars, which are composed of the CP4/He-weak stars and the He-rich stars \citep{preston74}. Relevant to the present investigation are the CP2 and the He-pec stars, which generally exhibit strong and globally organised magnetic fields and atmospheres selectively enriched in elements such as Si, Cr, Sr, or Eu. These objects are frequently grouped together under the term of magnetic chemically peculiar (mCP) stars, which, for convenience, we will adhere to in the present study.

Magnetic CP stars usually present a non-uniform surface distribution of chemical elements (chemical spots or belts). Flux is redistributed in these structures via line and continuum blanketing (e.g. \citealt{lanz96,shulyak10}), which results in strictly periodic light and spectral variations with the rotation period. After their bright prototype, photometrically variable mCP stars are also referred to as $\alpha^{2}$ Canum Venaticorum (ACV) variables. Observed amplitudes of the photometric variability are generally of the order of several hundredth of a magnitude in the optical region, occasionally reaching up to about 0.2\,mag ($V$). In the 'zoo' of variable stars, ACV variables constitute a rather obscure class and, in automated classification systems, are often combined with other rotational variables under generic types, such as 'ROT' (i.e. spotted stars not classified into a particular class), or get largely ignored.

Here we show that the 'upside-down CBH variables'\footnote{For convenience, in the text, these objects are hereafter referred to as ud-CBH variables.}, identified through the analysis of data from the Asteroid Terrestrial-impact Last Alert System (ATLAS) and proposed by \citet{heinze18} in the ATLAS variable star Data Release One (ATLAS DR1) catalogue as a potentially new class of variable stars, are, at least to a high percentage, made up of ACV variables with distinct double-wave light curves. Using suitable selection criteria, we identified 264 candidate ACV variables in the ATLAS DR1 catalogue. 62 of these stars could be confirmed as mCP stars (53 new discoveries) using spectra from the Large Sky Area Multi-Object Fiber Spectroscopic Telescope (LAMOST) and were classified on the MK system.

Data sources, motivation and the process of target selection are detailed in Section~\ref{datasources}. Spectral classifications are described and presented in Section \ref{spectral_classification}. The magnitude, period and space distributions of our sample stars are investigated in Section~\ref{space_distribution}, while Section \ref{parameters} explores the ages and masses of our sample stars. We conclude in Section~\ref{conclusions}.

\section{Data sources, motivation and target selection} \label{datasources}

\subsection{The Asteroid Terrestrial-impact Last Alert System (ATLAS)}

The Asteroid Terrestrial-impact Last Alert System (ATLAS) is a NASA-funded sky survey developed by the University of Hawaii with the main aim of discovering potentially hazardous near-Earth asteroids (NEAs). It currently consists of two telescopes, which are situated on Haleakala and Mauna Loa and equipped with f/2.0 0.5-m DFM custom Wright Schmidt telescopes and STA-1600 10.5x10.5k CCD detectors. The telescopes command a field of view with a diameter of 7\fdg4, and observations are procured through two broadband filters: the cyan filter ('$c$'; 420$-$650\,nm), which is employed during the two weeks surrounding the new Moon, and the orange filter ('$o$'; 560$-$820\,nm), which is used in lunar bright time. Both filters are linked via colour transformations to the Pan-STARRS $g$, $r$, and $i$ bands. More information on the ATLAS sky survey is found in \citet{tonry18}.


\subsection{The Large Sky Area Multi-Object Fiber Spectroscopic Telescope (LAMOST)} \label{LAMOST}

The LAMOST telescope \citep{lamost1,lamost2}, also called the Guo Shou Jing\footnote{Guo Shou Jing (1231$-$1316) was a Chinese astronomer, hydraulic engineer, mathematician, and politician of the Yuan Dynasty.} Telescope, is a reflecting Schmidt telescope located at the Xinglong Observatory in Beijing, China. It boasts an effective aperture of 3.6$-$4.9\,m and a field of view of 5$\degr$. Thanks to its unique design, LAMOST is able to take 4000 spectra in a single exposure with spectral resolution R\,$\sim$\,1800, limiting magnitude r\,$\sim$\,19\,mag and wavelength coverage from 3700\,\AA\ to 9000\,\AA. LAMOST is therefore particularly suited to survey large portions of the sky and is dedicated to a spectral survey of the entire available northern sky. LAMOST data products are released to the public in consecutive data releases and can be accessed via the LAMOST spectral archive.\footnote{\url{http://www.lamost.org}} With about 10 million stellar spectra contained in data release (DR) 6, the LAMOST archive constitutes a real treasure trove for researchers. The present study uses spectra from LAMOST DR4 \citep{DR4}.

\subsection{ATLAS variable star Data Release One (ATLAS DR1) and the 'upside-down CBH variables'}

Using data from the first two years of operation of the Haleakala telescope (observations up to the end of June 2017), \citet{heinze18} published the ATLAS DR1 catalogue. Analysing observations for approximately 142 million stars between declinations of about $-$30$^{\circ}$\,$\le$\,$\delta$\,$\le$\,+60$^{\circ}$ in the magnitude range of about 11\,<\,$r$\,<\,19, they produced a major catalogue containing $\sim$430,000 confirmed variable stars (among which are $\sim$300,000 new discoveries) and 4.7 million candidate variable stars. The observations consisted of four to six 30-s exposures of each field per night; the provided data sets boast a median time span of about 620\,d, the median number of observations is $N$\,=\,208 \citep{heinze18}. Subsequent data releases are planned, which are expected to greatly enhance the number of identified variables and available measurements.

\begin{figure}
	\includegraphics[width=\columnwidth,bb=0 0 500 400]{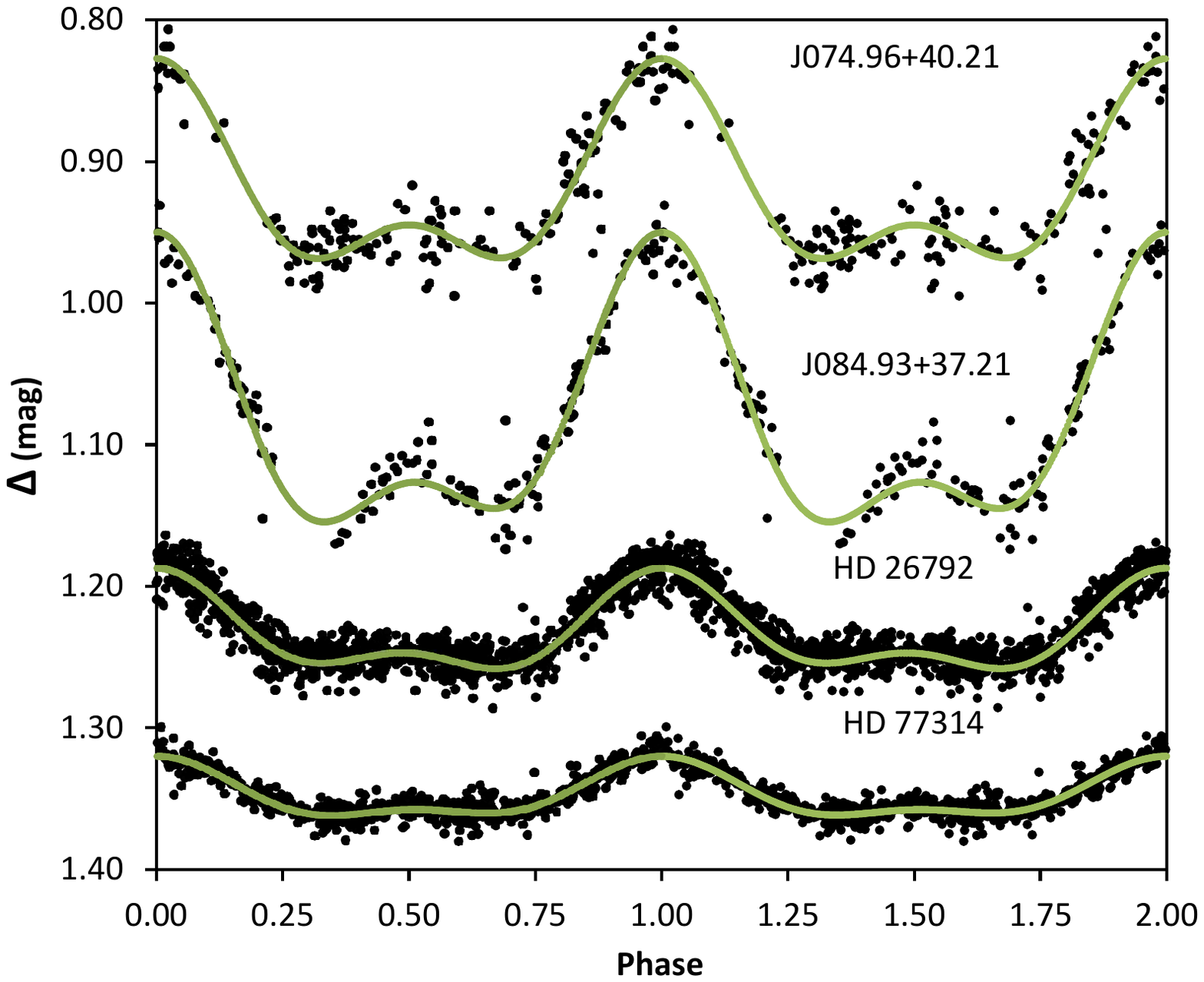}
    \caption{Phase plots of (from top to bottom) the 'upside-down CBH variables' ATO J074.9570+40.2063 and ATO J084.9321+37.2119, based on ATLAS data, and the known ACV variables HD 77314 and HD 26792, based on MASCARA data. The amplitude of the light variations appears reduced in the filterless MASCARA data.}
    \label{fig:LC_overview}
\end{figure}

\citet{heinze18} calculated an extensive set of 169 variability features for each object, 70 of which were employed as input for sorting the stars into broad, astrophysically meaningful variability classes based on light curve morphology, such as CBF = 'Close binary, full period'; CBH = 'Close binary, half period'; MIRA = 'High-amplitude, long-period red variable'; PULSE = 'Pulsating variable' etc. A complete overview of the variability classes is provided in Table 2 of \citet{heinze18}. 

In the context of the present study, the class of CBH = 'Close binary, half period' is of importance, which consists of close eclipsing binaries for which a period equal to half the true orbital period was selected by the employed period search algorithm. In analogy to this group of objects, and in want of a better expression, \citet{heinze18} coined the term ud-CBH variables to refer to a particular set of light curves that look almost exactly like the inverted light curves of CBH stars -- that is, instead of the minima (eclipses) in the light curves of CBH binaries, they show narrow symmetrical maxima (cf. Figure \ref{fig:LC_overview}, upper panels). About 70 such objects were manually identified, most of which were sorted into the PULSE class by the classification algorithm. However, the authors note that many more can probably be found in the final catalogue.

\citet{heinze18} briefly consider the astrophysical nature of the ud-CBH variables, which occupy a well-defined region in period-amplitude space (amplitudes of up to 0.2\,mag, period distribution peaks around 2\,d; cf. Figure 23 of \citealt{heinze18}). In particular, they rule out contact binaries (GCVS-type EW; \citealt{GCVS}) and binary stars with strong reflection effects (GCVS-type R) on the grounds of light curve morphology and argue against 'ordinary' rotating variables with classical starspots and binary systems with an accreting compact object because of the consistency of the observed light curves. In summary, the authors were unable to properly classify these stars and present them as a potentially new type of (pulsating?) variables, whose astrophysical nature is unclear.

\subsection{Motivation and target selection} \label{motivation_and_target_section}

The present study was conceived after we noticed the strong similarity of the light curves of the ud-CBH variables illustrated in \citealt{heinze18} (their Figure 23) to the light curves of a certain subset of ACV variables. Unlike stars with classical starspots, these spotted rotators show simple and remarkably consistent light curves that remain stable in shape and amplitude over decades and more \citep[e.g.][]{huemmerich18}. As has been expanded upon in Section \ref{introduction}, the light variations of ACV variables are due to the presence of chemical spots. They generally show simple light curves that can be well described by a sine wave and its first harmonic (e.g. \citealt{north84,mathys85,bernhard15b}). Depending on the inclination of the rotational axis and the number of spots that come into view during the rotation cycle, most ACVs show either single-wave or double-wave light curves \citep[e.g.][]{maitzen80,jagelka19}. In the case of two dominant spots of different sizes or photometric properties on opposing hemispheres of the star, a double-wave light curve with maxima of different heights is to be expected, in agreement with the light curves of the ud-CBH variables. This variability pattern is fairly unique among early-type stars and, according to our experience \citep{bernhard15b,bernhard15a,huemmerich16}, a tell-tale sign of ACV variables. According to the study of \citet{jagelka19}, roughly one third of ACV variables are expected to show double-wave light curves, with half of these stars exhibiting maxima of different height.

\begin{figure}
	\includegraphics[width=\columnwidth,bb=0 0 500 300]{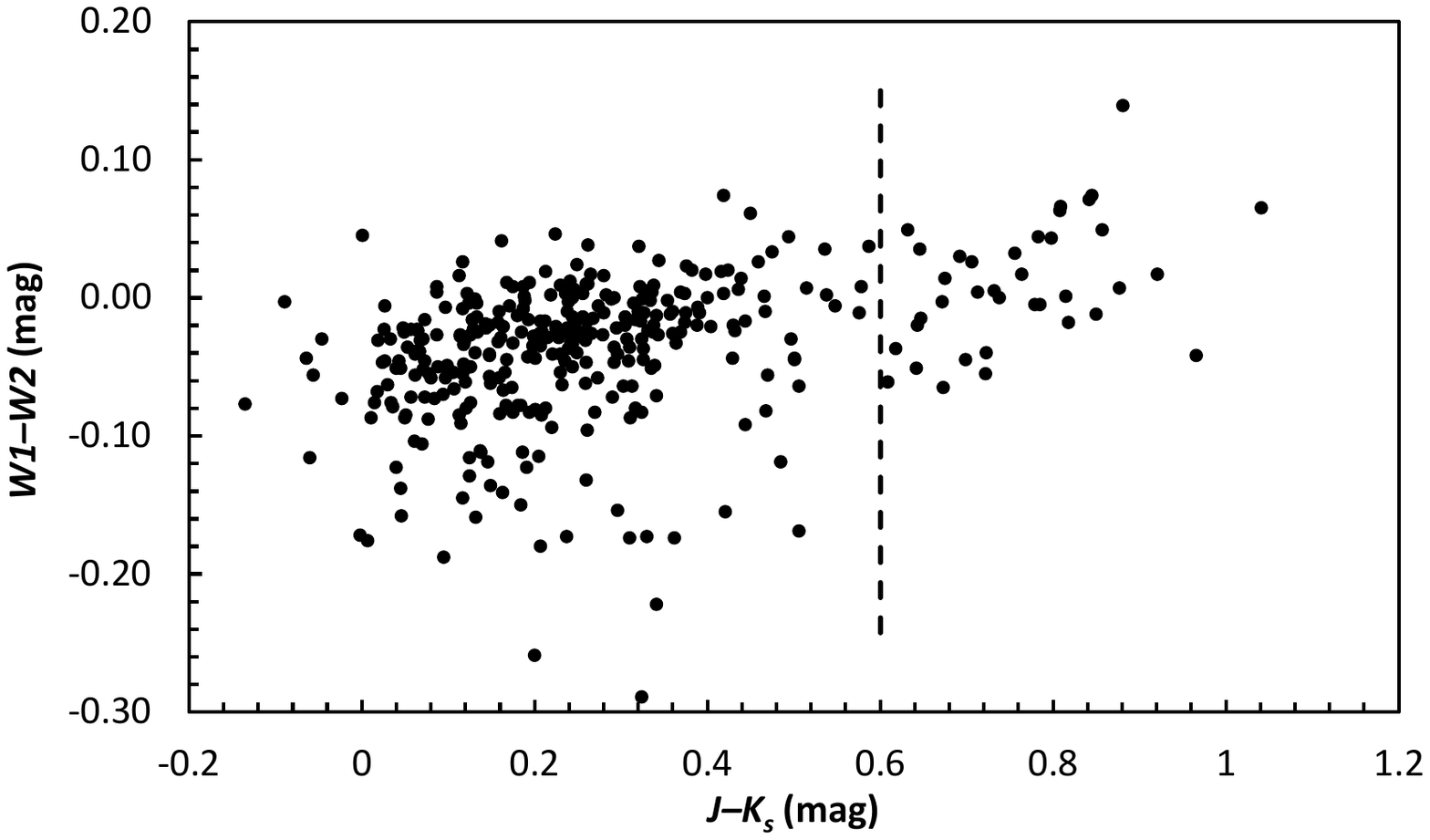}
    \caption{Near-/mid-infrared colour-colour diagram of all stars selected via the criteria outlined in Section \ref{motivation_and_target_section}. Note the 'red tail' of objects with $(J-K_{s})$\,$>$\,0.6\,mag.}
    \label{fig:colourcolourdiagram}
\end{figure}

Furthermore, the characteristics of the ud-CBH variables given in \citet{heinze18} -- low-amplitude ($<$0.2\,mag) variability;  periods in the range of about 1\,<\,$P(d)$\,<\,5, with a peak around 2\,d -- are in agreement with the majority of ACV variables, which mostly exhibit rotational periods in the above mentioned period range, with a well-known peak around 2\,d \citep[e.g.][]{RM09,bernhard15a}. A comparison of the light curves of two ud-CBH variables from the list of \citet{heinze18} to the light curves of two double-waved ACV variables taken from \citet{bernhard20} is provided in Figure \ref{fig:LC_overview}. While the amplitude of the light variations appears reduced in the filterless data from the Multi-site All-Sky CAmeRA (MASCARA; \citealt{MASCARA1,MASCARA2}), the general variability pattern is the same. As an initial test of our hypothesis, we investigated the LAMOST DR4 spectra of several ud-CBH stars, which confirmed our suspicion that these objects are indeed mCP (mostly CP2) stars.

For the final target selection process, several considerations had to be taken into account. As 'ud-CBH stars' is not one of the final adopted variability classes, these objects cannot be simply queried via the interface to the ATLAS DR1 catalogue.\footnote{More information on how to query the ALTAS DR1 catalogue is available at \url{https://archive.stsci.edu/prepds/atlas-var/}.} \citet{heinze18} comment that most ud-CBH variables were sorted into the category PULSE by the classification algorithm. Apart from that, we consider the CBH and SINE groups the most promising targets for discovering ACV variables among the remaining variability classes. The SINE group ('Pure sine was fit with small residuals') was chosen because most ACV stars (roughly 50\,\% of the sample of \citealt{jagelka19}) show symmetric single-wave light curves that can be well fitted by a sine wave. All other variability groups were ruled out as primary targets on grounds of the covered period and amplitude ranges, the occurrence of multiple periods, irregular variability, or, as ACV variables show remarkably stable light curves, the indication of large residuals to the sine fit (class NSINE = 'Noisy sine: pure sine was fit, but residuals are large or nonrandom').

On grounds of these considerations, and after some initial testing regarding the feasibility of our intended approach, we decided to investigate objects from the PULSE, CBH, and SINE groups with the following requirements: (i) 0.5\,<\,$P(d)$\,<\,10, (ii) peak-to-peak variability amplitude in the $c$ band of amp($c$)\,$<$\,0.3\,mag, and (iii) phase offset $\phi{_2}$\,$-$\,$\phi{_1}$ of the first two Fourier terms between 60$^{\circ}$ and 110$^{\circ}$. The imposed cuts form a compromise between an effective identification of mCP stars and the need to keep the sample down to a manageable size. The period (i) and phase offset (iii) requirements were enforced to exclude $\delta$ Scuti (GCVS-type DSCT) stars or other short period pulsators and W UMa variables. Item (iii) was chosen according to the results of \citet{heinze18}, who find that most ud-CBH stars have phase offsets near 90$^{\circ}$ (cf. in particular their Figure 23). We note that while there exist mCP stars with rotation periods in excess of the chosen limit, their incidence drops significantly towards longer rotational periods (e.g. \citealt{RM09}); more importantly, the number of suitable light curves in the ATLAS DR1 catalogue diminishes rapidly at $P$\,>\,10\,d. The peak-to-peak variability cut-off (ii) was set deliberately high in order not to per se exclude any high-amplitude ACV variables. Although most ACV stars show amplitudes smaller than 0.2\,mag ($V$), we noticed several stars with amplitudes approaching this limit in the $c$ band (cf. Figure \ref{fig:LC_overview}).

Frequency spectra of all objects selected via these criteria were calculated using \textsc{Period04} \citep{period04}. Light curves and frequency spectra were then visually inspected. Obviously multiperiodic objects, such as slowly pulsating B (GCVS-type SPB) stars and $\gamma$ Doradus (GCVS-type GDOR) variables, were discarded; mono-periodic objects (that is, objects with frequency spectra indicative of a single significant frequency or a single significant frequency plus corresponding harmonics) with stable light curves typical of ACV variables were chosen for further consideration.

\begin{figure*}
	\includegraphics[width=2.0\columnwidth,bb=0 0 600 400]{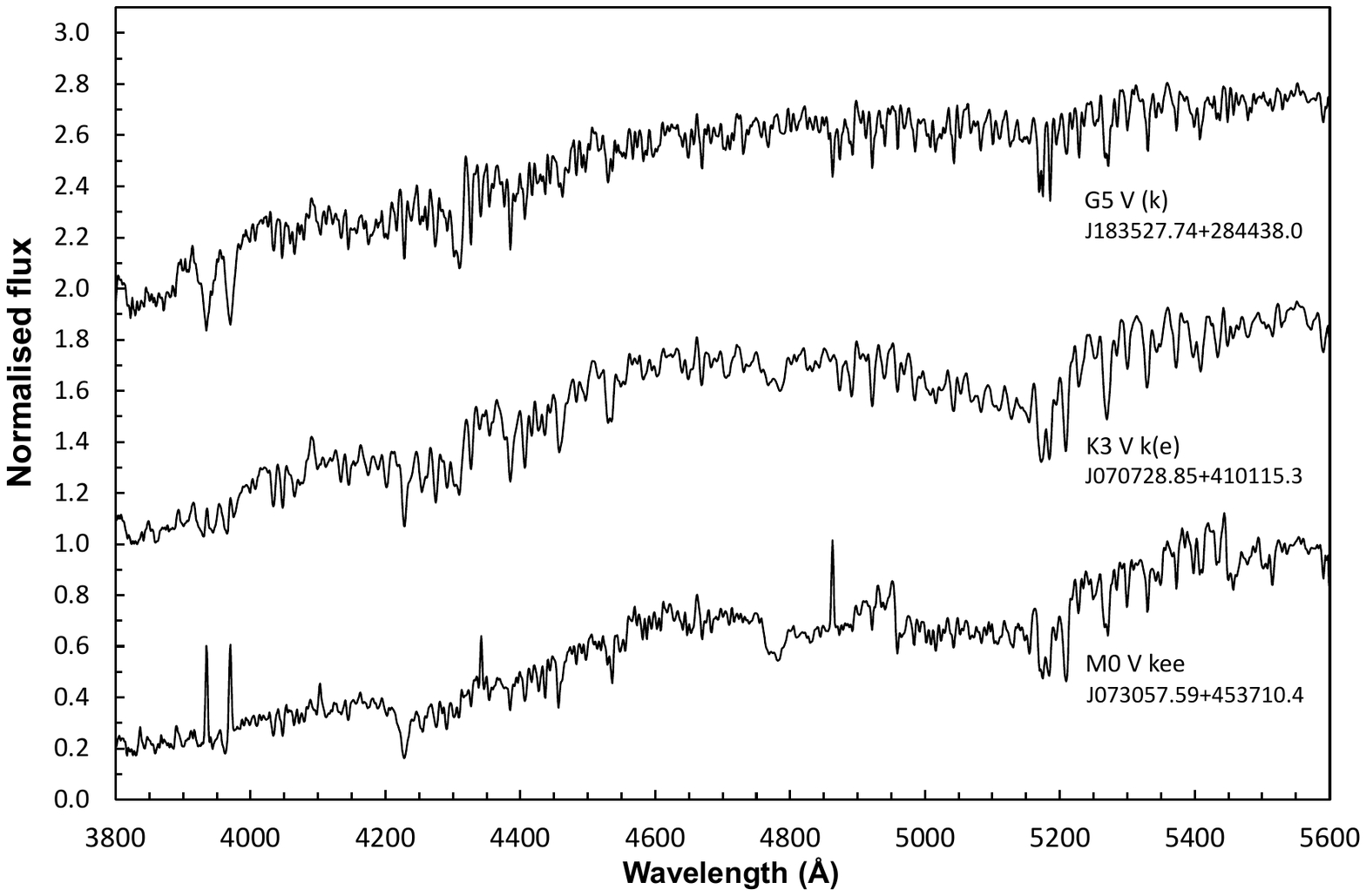}
    \caption{Spectra of late-type rotational variables from the 'red tail' with $(J-K_{s})$\,$>$\,0.6\,mag, showing signs of increased chromospheric activity. Identifiers and spectral types (derived in this study) are indicated in the plot. The spectrum of J070728.85+410115 shows emission in the H alpha line (not shown in the plot).}
    \label{fig:showcase_red_stars}
\end{figure*}

Although spot configurations in late-type stars are much more unstable, we expect contamination by late-type rotational variables such as BY Draconis (GCVS-type BY) or RS Canum Venaticorum (GCVS-type RS) stars that, at least in the investigated time-span, did not show significant changes in spot sizes and distribution and therefore exhibited appreciably stable light curves. For this reason, as an initial check, the remaining objects were plotted in a near-/mid-infrared colour-colour diagram (Figure \ref{fig:colourcolourdiagram}) in order to identify red stars. To this end, we employed photometry from the 2MASS \citep{2MASS} and WISE \citep{WISE} catalogues. Considering the faintness of the investigated objects and the unknown amount of reddening in the line-of-sight, this will only have a limited predictive value as regards the intrinsic colours of the stars. However, there is an obvious red tail of objects with ($J-K_{s}$)\,$>$\,0.6\,mag, which likely corresponds to late-type rotational variables. For several of these objects, LAMOST DR4 spectra are available, which indeed confirmed this assumption. Three exemplary spectra are shown in Figure \ref{fig:showcase_red_stars}. As expected, all objects show signs of increased chromospheric activity, which is typical of late-type rotational variables. We therefore felt justified in excluding objects with ($J-K_{s}$)\,$>$\,0.6\,mag from further consideration.

264 stars passed the imposed selection criteria, which, for convenience, are hereafter referred to as the 'final sample'. The full list of parameters for these objects is provided in Table \ref{table_master_part1} in the Appendix (Section \ref{essentialdata}). A sample page illustrating the light curves of the first 16 stars from our sample is also provided in the Appendix (Figure \ref{LC_sample_page}). The full set of light curves is obtainable online.

\section{Spectral classification} \label{spectral_classification}

The final sample was cross-matched with the LAMOST DR4 VizieR online catalogue\footnote{\url{http://cdsarc.u-strasbg.fr/viz-bin/cat/V/153}} \citep{DR4}, requiring S/N\,$\ge$\,25 in the Sloan $g$ band. In this manner, spectra for 62 stars could be procured. If more than one spectrum was available for a single object, only the spectrum with the highest $g$ band S/N was considered.

Spectral classification was performed in the framework of the refined MK classification system following the methodology outlined in \citet{gray87,gray89a,gray89b,gray94} and \citet{gray09}. For a precise classification and to identify peculiarities, the blue-violet (3800$-$4600\,\AA) spectral region was compared visually to, and overlaid with, MK standard star spectra and a set of synthetic spectra.

The standard star spectra were taken from the \textit{libr18} and \textit{libnor36} libraries that are distributed with the MKCLASS code\footnote{\url{http://www.appstate.edu/~grayro/mkclass/}}, a computer program written to classify stellar spectra on the MK system \citep{gray14}, and the \textit{liblamost} library, which is a preliminary set of suitable LAMOST spectra compiled by \citet{huemmerich20}. The synthetic spectra were calculated using the program SPECTRUM\footnote{\url{http://www.appstate.edu/~grayro/spectrum/spectrum.html}} \citep{SPECTRUM} and ATLAS9 model atmospheres \citep{ATLAS9}, assuming log g = 4.0, [M/H] = 0.0 and a microturbulent velocity of 2 km\,s$^{-1}$. The spectra were then subsequently smoothed to a resolution of 3.0\,\AA\ to provide a good match to the LAMOST spectra.

\begin{figure*}
	\includegraphics[width=2.1\columnwidth,bb=0 0 650 400]{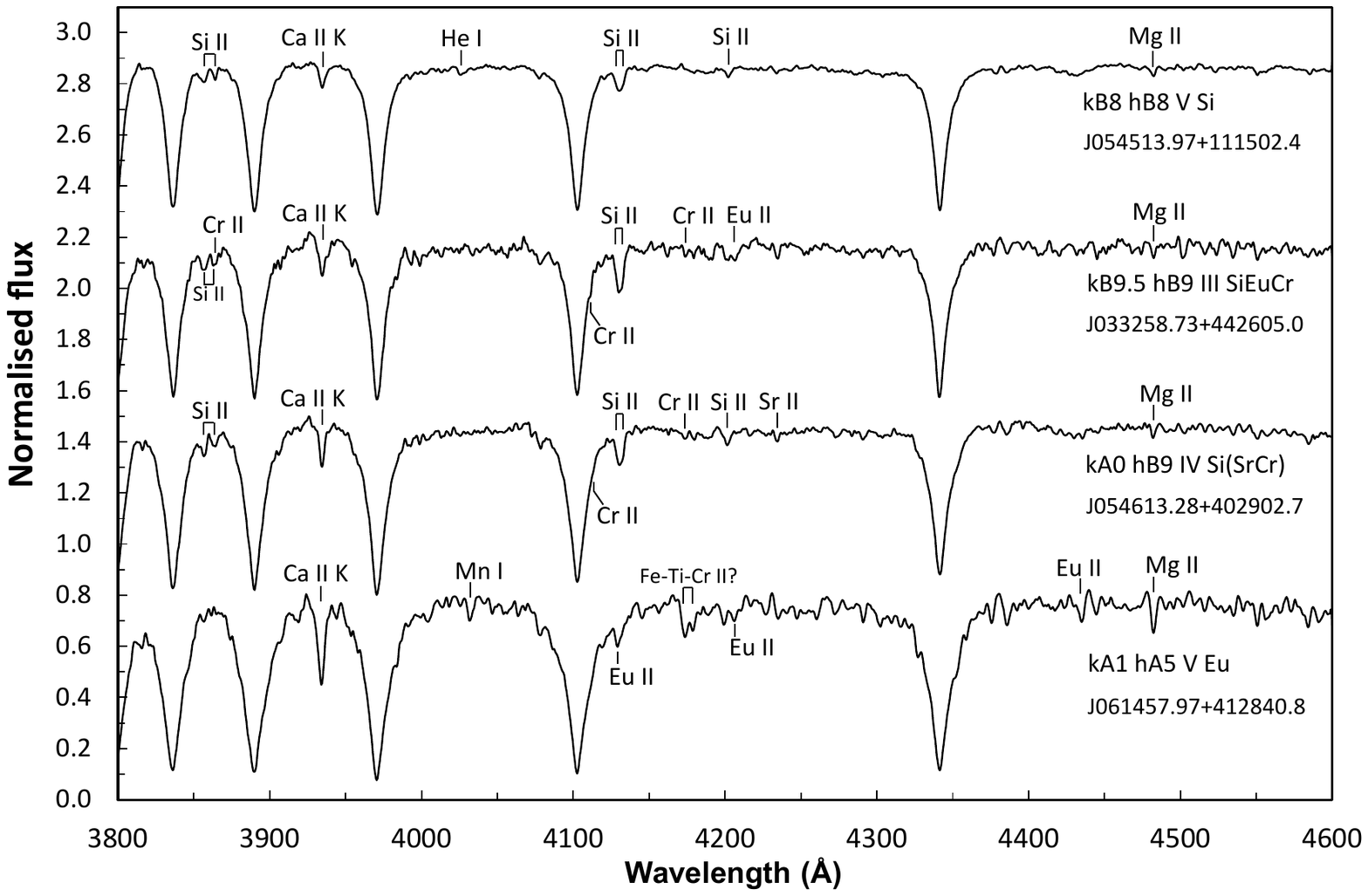}
    \caption{Blue-violet region of the LAMOST DR4 spectra of (from top to bottom) the CP2 stars LAMOST J054513.97+111502.4 (ATO J086.3082+11.2506), LAMOST J033258.73+442605.0 (ATO J053.2447+44.4347), LAMOST J054613.28+402902.7 (ATO J086.5553+40.4840), and LAMOST J061457.97+412840.8 (ATO J093.7415+41.4780). Some prominent lines of interest are identified.}
    \label{fig:showcase_ATLAS}
\end{figure*}

mCP stars exhibit several peculiarities related to traditional classification criteria, which need to be taken into account when assigning a spectral type to these objects. They often show weak, unusually broad or otherwise peculiar \ion{Ca}{ii} K line profiles, weak \ion{Mg}{ii} 4481\,\AA\ lines and are markedly He deficient \citep{gray09,ghazaryan18}. The latter characteristic resulted in the classification of many of the hotter (late B-type) mCP stars as A-type stars because of the lack of, or the occurrence of only very weak, \ion{He}{i} lines in their spectra \citep{gray09}. Furthermore, mCP stars generally show enhanced and peculiar metallic lines. In general, therefore, the hydrogen-line profile is the most accurate indicator of the actual temperature of these objects \citep{gray09}.

We have determined spectral types based on the \ion{Ca}{ii} K line strength (the k-line type) and the hydrogen-line profile (the h-line type) \citep{osawa65}. Spectral types based on the strength of the \ion{He}{i} lines were not determined for all stars but only in the case of the He-weak/CP4 stars and where they were unambiguously visible in the spectra. Most late B-type stars in our sample appear noticeably He weak.

The metallic lines of most mCP stars are so peculiar that they cannot be used for luminosity classification. Luminosity types, therefore, were based on the wings of the hydrogen lines. In this respect, it is important to note that a considerable number of mCP stars shows narrow or unusual hydrogen-line profiles \citep[e.g.][]{faraggiana87,gray94,ghazaryan18} which are best matched by the profiles of high-luminosity objects like giants or even bright giants (luminosity classes III and II, respectively). This also holds true for many stars of our sample, although, apart from the hydrogen-line profile, other high-luminosity indicators are mostly absent from their spectra. It is well known that the vast majority of mCP stars are in fact main-sequence objects (\citealt{netopil17} and references therein) and our results from the colour-magnitude diagram (CMD) (Section \ref{parameters}) fully support this finding. Non-LTE effects may be at the root of this obvious discrepancy between the luminosities derived from the hydrogen-line profiles and other indicators \citep{gray94}. Perhaps the influence of the strong magnetic fields results in unusual atmospheric structures; however, this remains as yet unsolved.

Main chemical peculiarities are given in order of importance. Parantheses and a double pair of parantheses have been employed to indicate that a peculiarity is, respectively, weakly or only marginally present \citep{gray87,gray89a,gray89b,gray94}. Likewise, the notation ``metallic lines'' is used to indicate a faint background of metallic lines due to other atomic species than iron, titanium, or their ions \citep{gray87}. This notation may also appear in parantheses or a double pair of parantheses.

For several objects, only spectra of low S/N (S/N < 40) are available. Spectral types based on these spectra should be regarded as estimates only. For instance, the difference between the hydrogen-line profiles of a B8 III star and a B5 V star gets blurred in low S/N spectra \citep{gray94}. As mCP stars often show weak He lines, this leads to corresponding uncertainties in the spectral classification.

Example spectra are provided in Figure \ref{fig:showcase_ATLAS}; final spectral types and remarks are presented in Table \ref{table_SpT}. We also investigated the 4800$-$5600\,\AA\ spectral region to check for the presence and morphology of the flux depression around 5200\,\AA, which is a characteristic of mCP stars \citep{kodaira69,kupka03,khan07} and readily visible at LAMOST resolution \citep{huemmerich20}. Unless indicated otherwise in the remarks, all objects show a clearly visible 5200\,\AA\ feature in their spectra, as expected.

Nine of the spectroscopically confirmed mCP stars are contained in the sample of \citet{huemmerich20}; the remaining 54 stars are here presented as mCP stars for the first time. Our classifications agree well with the automatically derived classifications from the aforementioned study. \citet{huemmerich20} did not specify k-line types for all objects; however, the h-line types, the most important temperature-type criterion in mCP stars, agree very well: the differences amount to 0.0 spectral subclasses in the case of six objects, 0.5 spectral subclasses in the case of two stars, and 2.0 spectral subclasses in the case of one star (one spectral subclass being the difference between, for instance, A0 and A1). The main spectral peculiarities have been correctly identified by \citet{huemmerich20}; using manual classification, we can in some cases add weak peculiarities that have been missed by their code and, in two cases, are able to resolve the generic 'bl4077' (strong 4077\,\AA\ blend) and 'bl4130' (strong 4130\,\AA\ blend) types into their contributing ions. For instance, in the case of J072338.12+003848.5, 'bl4077' resolves into \ion{Si}{II} 4076\,\AA\ and \ion{Sr}{II} 4077\,\AA, and 'bl4130' into \ion{Si}{II} 4128$-$30\,\AA\ and \ion{Eu}{II} 4130\,\AA. Classifications from \citet{huemmerich20}, where available, are also provided in Table \ref{table_SpT}.

\begin{table*}
\caption{Spectral types of the mCP stars identified in the present study. The columns denote: (1) ATLAS object identifier. (2) LAMOST identifier. (3) Sloan $g$ band S/N ratio of the analysed spectrum. (4) Spectral type, as derived in this study. (5) Remark. 'H20' denotes spectral types from the study of \citet{huemmerich20}.}
\label{table_SpT}
\begin{center}
\begin{adjustbox}{max width=0.9\textwidth}
\begin{tabular}{lllll}
\hline
\hline
(1) & (2) & (3) & (4) & (5) \\
\hline
ID\_ATO	&	ID\_LAMOST	&	S/N\,$g$	&	SpT\_final	&	remark	\\
\hline
\hline
J046.2665+57.4649	&	J030503.98+572753.7	&	33	&	kA0 hB7 HeB8 V Si(Sr:)	&	no 5200\,\AA\ flux depression visible	\\
J046.9585+53.9453	&	J030750.05+535643.2	&	98	&	kA0 hA0: III-IV (Si)	&	$\lambda$4481 weak	\\
J047.5021+53.3159	&	J031000.50+531857.3	&	30	&	kA1 hA3 III SiCrEu	&		\\
J050.9303+44.3793	&	J032343.27+442245.4	&	108	&	kA0 hA0 II EuSi	&	$\lambda$4481 weak; H20: A0 Ib-II EuSi	\\
J053.2447+44.4347	&	J033258.73+442605.0	&	69	&	kB9.5 hB9 III SiEuCr	&	$\lambda\lambda$4172-9 strong	\\
J057.4910+58.7724	&	J034957.85+584620.8	&	57	&	kA0 hB9.5 III Si(Cr)Eu	&	$\lambda$4481 weak	\\
J060.7275+46.2448	&	J040254.59+461441.4	&	45	&	kA1 hB9: II: SiEuSr	&	$\lambda$4481 weak	\\
J060.7480+45.3172	&	J040259.53+451902.0	&	113	&	kA1 hA1 III SiEu((Cr))	&	$\lambda$4481 weak	\\
J063.5805+46.9075	&	J041419.32+465427.1	&	42	&	knA1 hA5 III EuCrSr(Si)	&	$\lambda$4481 weak	\\
J064.2611+46.2848	&	J041702.66+461705.5	&	65	&	kA0 hA0 III-IV Si	&	$\lambda$4481 weak	\\
J065.7038+47.6938	&	J042248.93+474138.0	&	308	&	kA0 hA0 II EuSiCr	&	$\lambda$4481 weak; H20: A0 II Eu	\\
J066.9854+44.3739	&	J042756.51+442226.2	&	67	&	kB9.5 hB8 IV Si	&	$\lambda$4481 weak	\\
J067.4887+38.9535	&	J042957.29+385712.5	&	58	&	knB9 hB5 V Si	&		\\
J072.5642+39.5294	&	J045015.40+393146.0	&	78	&	kA1 hA5 III SiCr	&		\\
J074.4268+43.8711	&	J045742.44+435216.0	&	30	&	kA0 hA5 III-IV SiEuSr	&	$\lambda$4481 weak	\\
J074.9570+40.2063	&	J045949.68+401222.8	&	83	&	kA0 hB9 III-IV SiEu	&	$\lambda$4481 broad and shallow	\\
J074.9889+42.2979	&	J045957.34+421752.6	&	31	&	kA0 hB9 III SiEuCr	&		\\
J076.4023+45.6101	&	J050536.57+453636.6	&	59	&	kA1 hB9 III EuSi	&	Eu very strong; $\lambda$4481 broad and shallow	\\
J077.0867+36.5911	&	J050820.81+363528.0	&	41	&	kB9.5 hB8 V (Si) ((metallic lines))	&	very weak 5200\,\AA\ flux depression	\\
J081.9629+42.4325	&	J052751.11+422557.3	&	92	&	kB9.5 hB9.5 III Si	&	$\lambda$4481 weak	\\
J082.5579+33.1925	&	J053013.91+331133.3	&	31	&	kA0 hB9: III: SiCr:Eu:	&	$\lambda$4481 weak	\\
J082.6115+29.6992	&	J053026.78+294157.2	&	84	&	kA0 hB9 III Si	&	$\lambda$4481 weak	\\
J082.6322+21.1693	&	J053031.74+211009.4	&	103	&	kB9 hB7 HeB9 V ((metallic lines))	&		\\
J084.9321+37.2119	&	J053943.72+371243.0	&	59	&	knA0 hB9 V SiEu((Cr))	&	$\lambda$4481 broad and shallow	\\
J086.1147+31.8873	&	J054427.54+315314.3	&	28	&	kA1 hB9 IV Si:Cr:Eu:	&		\\
J086.1972+24.3919	&	J054447.34+242331.1	&	52	&	kB9 hB7 HeB8 V ((Si)) (metallic lines)	&		\\
J086.3082+11.2506	&	J054513.97+111502.4	&	232	&	kB8 hB8 V Si	&	$\lambda$4481 weak	\\
J086.4762+23.3391	&	J054554.30+232021.0	&	47	&	kA0 hB9 III SrCrEuSr	&	$\lambda$4481 slightly weak	\\
J086.5553+40.4840	&	J054613.28+402902.7	&	98	&	kA0 hB9 IV Si(SrCr)	&	$\lambda$4481 weak	\\
J087.1757+29.1833	&	J054842.18+291059.9	&	35	&	kA0 hB9: V: SiEuCr	&	$\lambda$4481 weak	\\
J087.2218+29.4548	&	J054853.23+292717.3	&	26	&	kB9 hA0: III: SiCrEu	&	$\lambda$4481 weak	\\
J088.2315+38.6528	&	J055255.56+383910.1	&	47	&	kB9 hB7 HeB9 V Si (metallic lines)	&	$\lambda$4481 weak	\\
J090.0237+21.9017	&	J060005.70+215406.3	&	40	&	kA0 hB9 V CrSiEu	&	\\
J090.7455+31.0548	&	J060258.92+310317.1	&	37	&	kB9 hB9: II: SiCrSr	&	$\lambda$4481 weak	\\
J091.1746+24.8696	&	J060441.92+245210.8	&	26	&	kB9: hB9: III: Si	&	$\lambda$4481 weak	\\
J091.9664+24.5688	&	J060751.94+243407.9	&	95	&	kB9 hB5 HeB9 IV (Si) (metallic lines)	&		\\
J093.7415+41.4780	&	J061457.97+412840.8	&	112	&	kA1 hA5 V Eu	&	$\lambda\lambda$4172-9 strong	\\
J093.9930+13.8659	&	J061558.32+135157.3	&	148	&	kB9 hB5 HeB9 V (SiSr)Ti?	&	H20: B5 III-IV bl4130 (He-wk)	\\
J094.3323+15.5243	&	J061719.74+153127.5	&	86	&	kB9.5 hB7 HeB9 V Si	&		\\
J094.7507+19.5193	&	J061900.18+193109.6	&	76	&	knA0 hA0: II-III SiEuCr	&	$\lambda$4481 weak	\\
J095.0108+01.8715	&	J062002.59+015217.4	&	41	&	kA0 hB9 IV SiCrSrEu	&	$\lambda$4481 weak	\\
J095.4959+23.6564	&	J062159.02+233923.1	&	42	&	knB9 hB8 HeB9 V SiCr	&		\\
J096.8934+14.9451	&	J062734.41+145642.6	&	37	&	kA1 hA3: III: Cr:Si:	&	$\lambda$4481 weak	\\
J097.1763+16.4813	&	J062842.31+162852.9	&	168	&	kB9.5 hB9.5 III EuSi((Cr))	&	$\lambda$4481 weak; H20: B9.5 II-III EuSi	\\
J097.4242+24.5808	&	J062941.80+243450.9	&	155	&	kB9 hB7 HeB8 V 	&		\\
J097.6873+03.0070	&	J063044.96+030025.3	&	61	&	kB9 hB7 HeB9: V 	&		\\
J098.5904+17.0987	&	J063421.70+170555.6	&	83	&	knA0 hB9 III Si	&	$\lambda$4481 weak	\\
J098.6389+19.1051	&	J063433.34+190618.6	&	26	&	kA0 hB8 IV SiCr:Eu:	&	$\lambda$4481 weak	\\
J099.0745+18.8221	&	J063617.89+184919.6	&	28	&	kB9 hB9 IV SiEuCr	&		\\
J099.6853+17.3396	&	J063844.48+172022.5	&	162	&	kA0 hB9 III Si(Sr)((Cr))	&	$\lambda$4481 weak; H20: B9 III-IV Si	\\
J100.3063+00.9026	&	J064113.52+005409.4	&	43	&	kA0 hB8 IV SiCr	&	$\lambda$4481 weak	\\
J101.1743+21.6307	&	J064441.84+213750.5	&	132	&	kA0 hA0 II EuSi((Cr))	&	$\lambda$4481 weak; H20: A0 II EuSi	\\
J101.3758+01.7599	&	J064530.19+014535.8	&	97	&	kB9 hB8 V SiEu((Cr))	&	$\lambda$4481 weak	\\
J103.3079+10.5510	&	J065313.89+103303.3	&	119	&	kA0 hA0 III EuSi	&	$\lambda$4481 weak; H20: kA0hA2mA5 Eu	\\
J103.8576-00.4954	&	J065525.82-002943.4	&	71	&	kA0 hB5 V (metallic lines)	&		\\
J105.0327+06.1169	&	J070007.86+060701.0	&	50	&	kA1 hA1 II: EuCrSi	&	$\lambda$4481 weak	\\
J106.4471-03.7806	&	J070547.31-034650.2	&	80	&	kB9.5 hB9 III: (metallic lines)	&	\\
J106.5580+18.8987	&	J070613.94+185355.6	&	63	&	kA0 hB9 III Si	&	H20: B9.5 II-III Si	\\
J110.2675-03.2520	&	J072104.20-031507.4	&	142	&	kA0 hB9 III SiSrEu	&	$\lambda$4481 weak; H20: B9.5 II-III bl4077 bl4130	\\
J110.9088+00.6468	&	J072338.12+003848.5	&	94	&	knB9 hB9 IV-V SiSrCrEu	&	$\lambda$4481 weak; very broad Ca II K line blend	\\
J306.9801+45.3863	&	J202755.24+452311.0	&	99	&	kA0 hB9 IV Si(EuSr)	&	$\lambda$4481 weak	\\
J350.3833+56.4514	&	J232132.00+562705.2	&	57	&	kA0 hA3 III EuSi	&	$\lambda$4481 weak	\\
\hline
\end{tabular}                 
\end{adjustbox}
\end{center}                                                    \end{table*}

\section{Magnitude, period and space distribution} \label{space_distribution}

\begin{figure}
        \includegraphics[width=\columnwidth,bb=0 0 550 400]{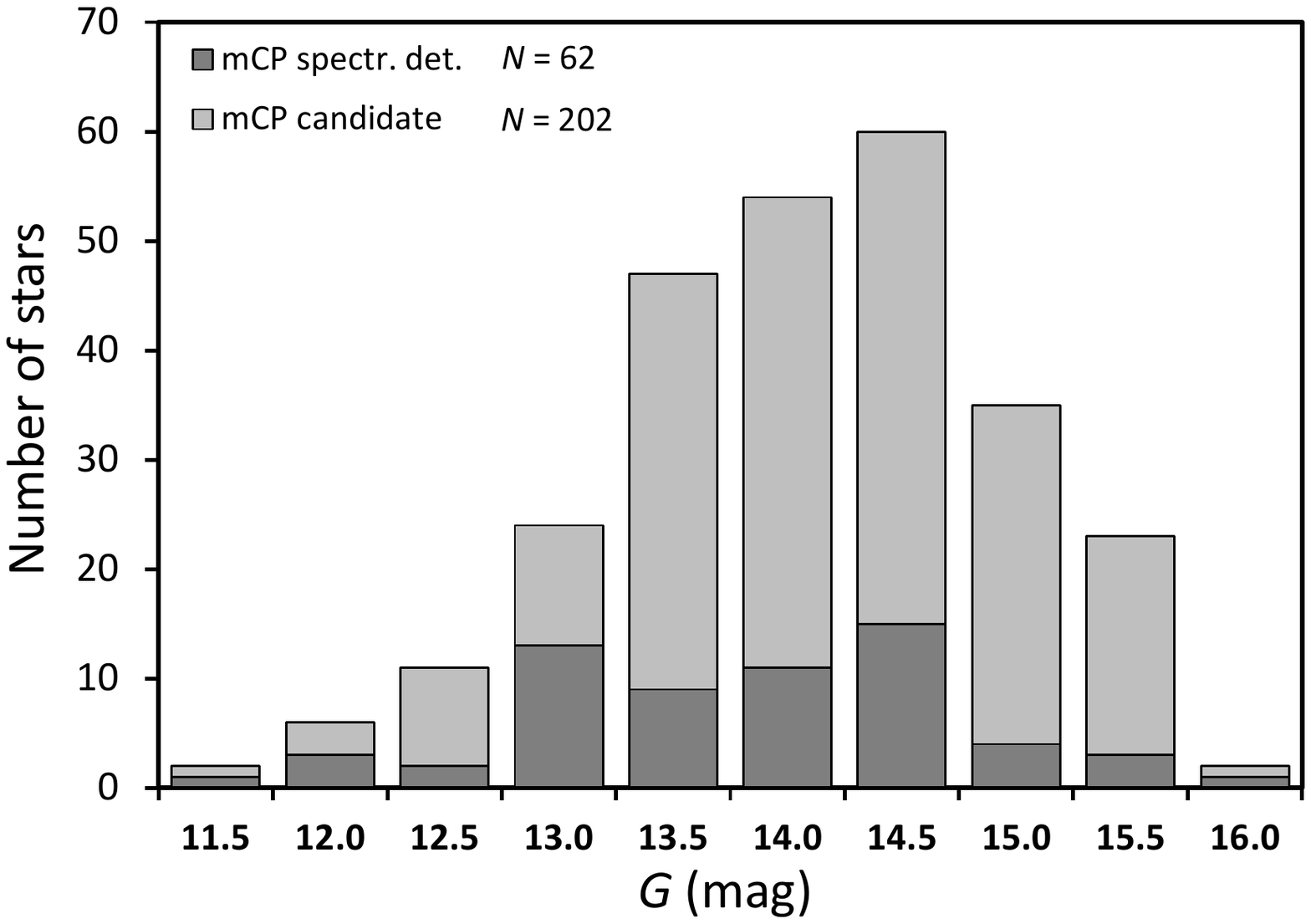}
        \includegraphics[width=\columnwidth,bb=0 0 550 400]{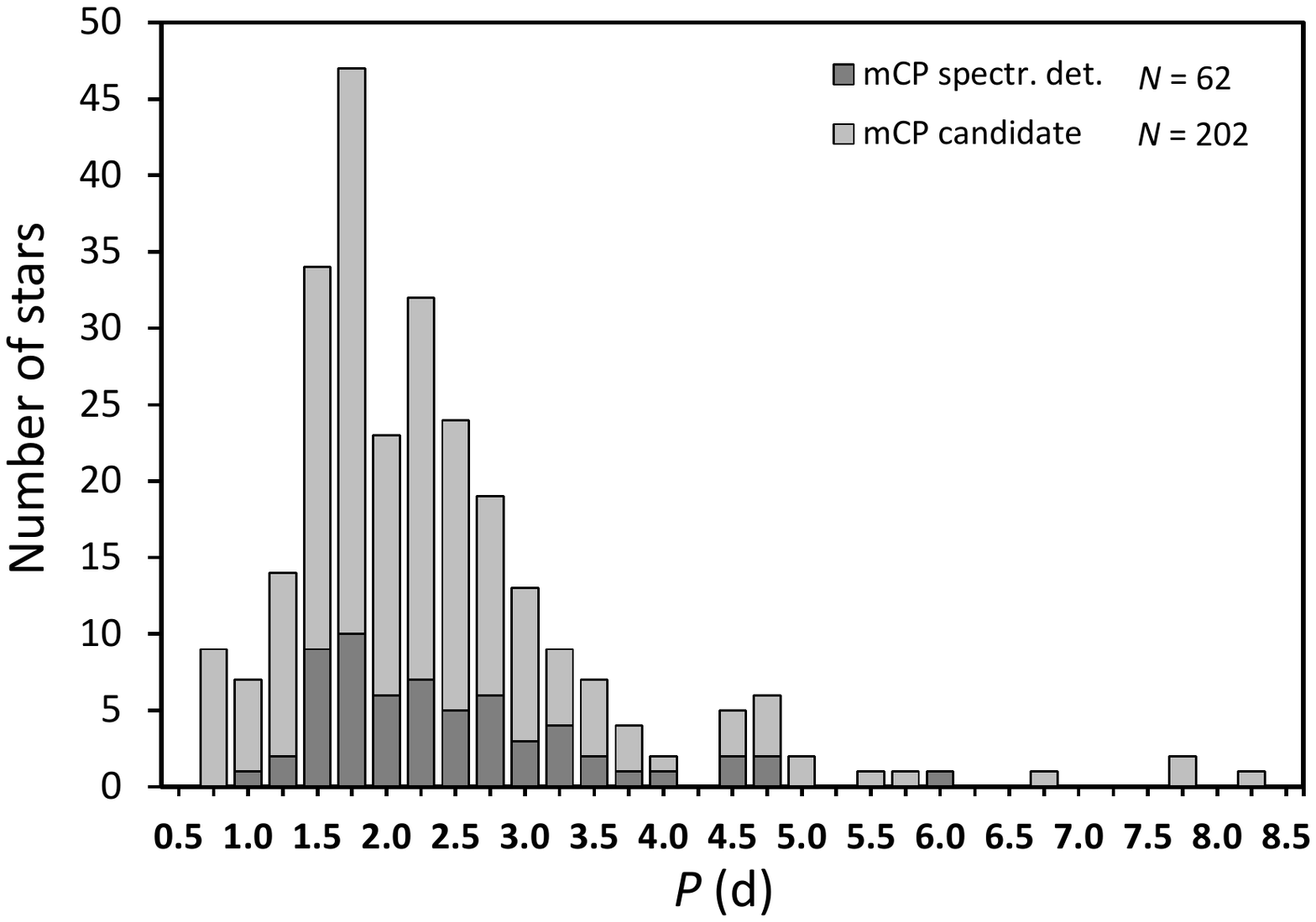}
    \caption{Histograms of the $G$ magnitudes (upper panel) and photometric periods (lower panel) of the spectroscopically confirmed mCP stars ($N$\,=\,62) and the mCP star candidates ($N$\,=\,202) of our final sample.}
		\label{histograms}
\end{figure}

The upper panel of Figure \ref{histograms} illustrates the histogram of the $G$ magnitudes of the spectroscopically confirmed mCP stars ($N$\,=\,62) and the mCP star candidates ($N$\,=\,202) of our final sample. The magnitude distribution peaks around 14th magnitude. For comparison, the stars in the catalogues of \citet{RM09} and \citet{huemmerich20} peak at 9th magnitude and between 11th and 12th magnitude, respectively. With this work, we therefore extend the search for new mCP stars to even fainter magnitudes. 

The period histogram is presented in the lower panel of Figure~\ref{histograms}. All periods were extracted from \citet{heinze18} and are presented, together with other parameters, in Table \ref{table_master_part1}. The period distribution is fully in agreement with the observed rotational periods of mCP stars, which generally cluster at around a period value of 2 days \citep[e.g.][]{RM09,bernhard15a}. We recall the here imposed period cut-off (0.5\,<\,$P(d)$\,<\,10), which was chosen to exclude short-period pulsators and because the number of suitable light curves in the ATLAS DR1 catalogue diminishes rapidly at $P$\,>\,10\,d (cf. Section \ref{motivation_and_target_section}). Any mCP stars with rotation periods longer than 10 days will therefore have been missed by our approach.

The coordinates in the Galactic [XYZ] plane were obtained via the conversion of spherical Galactic coordinates (latitude and longitude) to Cartesian coordinates using the distance $d$ from \citet{2021AJ....161..147B}. Only objects with absolute parallax errors less than 25\,\% ($N$\,=\,257) were considered. The positive X-axis is towards the Galactic centre, the positive Y-axis is in the direction of the Galactic rotation, and the positive Z-axis is towards the north Galactic pole. 

Our results are shown in the Appendix in Figure \ref{map_3D}. Most sample stars classify as either members of the thin disk (scale height of 350\,pc) or the thick disk (1200\,pc). The corresponding scale heights were adopted from \citet{2001MNRAS.322..426O} and \citet{2017MNRAS.470.2113A}. Interestingly, four stars (ATO J265.2024+10.6314, ATO J072.9129-07.0638, ATO J005.5419+15.5369, and ATO J310.1614-15.1051) are situated more than 1200\,pc above or below the [XY] plane and thus qualify as members of the halo population. This is of considerable interest as only very few halo CP2 stars are known \citep{huemmerich20}.

\section{Mass and age distribution} \label{parameters}

For generating a colour-magnitude diagram (CMD), we employed the homogeneous Gaia DR2 photometry of \citet{2018A&A...616A..17A} and the photogeometric distances of \citet{2021AJ....161..147B}, which are based on the Gaia EDR3 \citep{2021A&A...649A...3R}. As most of our sample stars are members of the Galactic disk and situated within a mean distance of 2500\,pc from the Sun, interstellar reddening is significant and needs to be taken into account. To estimate the absorption in the line-of-sight, we interpolated the reddening map of \citet{2018MNRAS.478..651G} using the Gaia EDR3 photogeometric distances. Reddening values were transformed using the relations
\begin{equation}
E(B - V) = 0.76E(BP - RP) = 0.40A_G,
\end{equation}
which take into account the coefficients listed in \citet{2018MNRAS.478..651G}.

The Gaia DR2 photometry was corrected as suggested by \citet{2018A&A...619A.180M}. The PARSEC isochrones \citep{2012MNRAS.427..127B} for solar metallicity [Z]\,=\,0.02\,dex and the filter curves from \citet{2018A&A...619A.180M}
were used for further analysis.
For the estimation of the stellar mass and logarithmic ages, the Stellar Isochrone Fitting Tool\footnote{\url{https://github.com/Johaney-s/StIFT}} was used, which searches for the best-fit evolutionary track within a model grid box. The four grid points closest to the input value are determined, from which the output parameters are derived by bilinear interpolation following the methodology described by \citet{2010MNRAS.401..695M}.

\begin{figure}
	\includegraphics[width=\columnwidth]{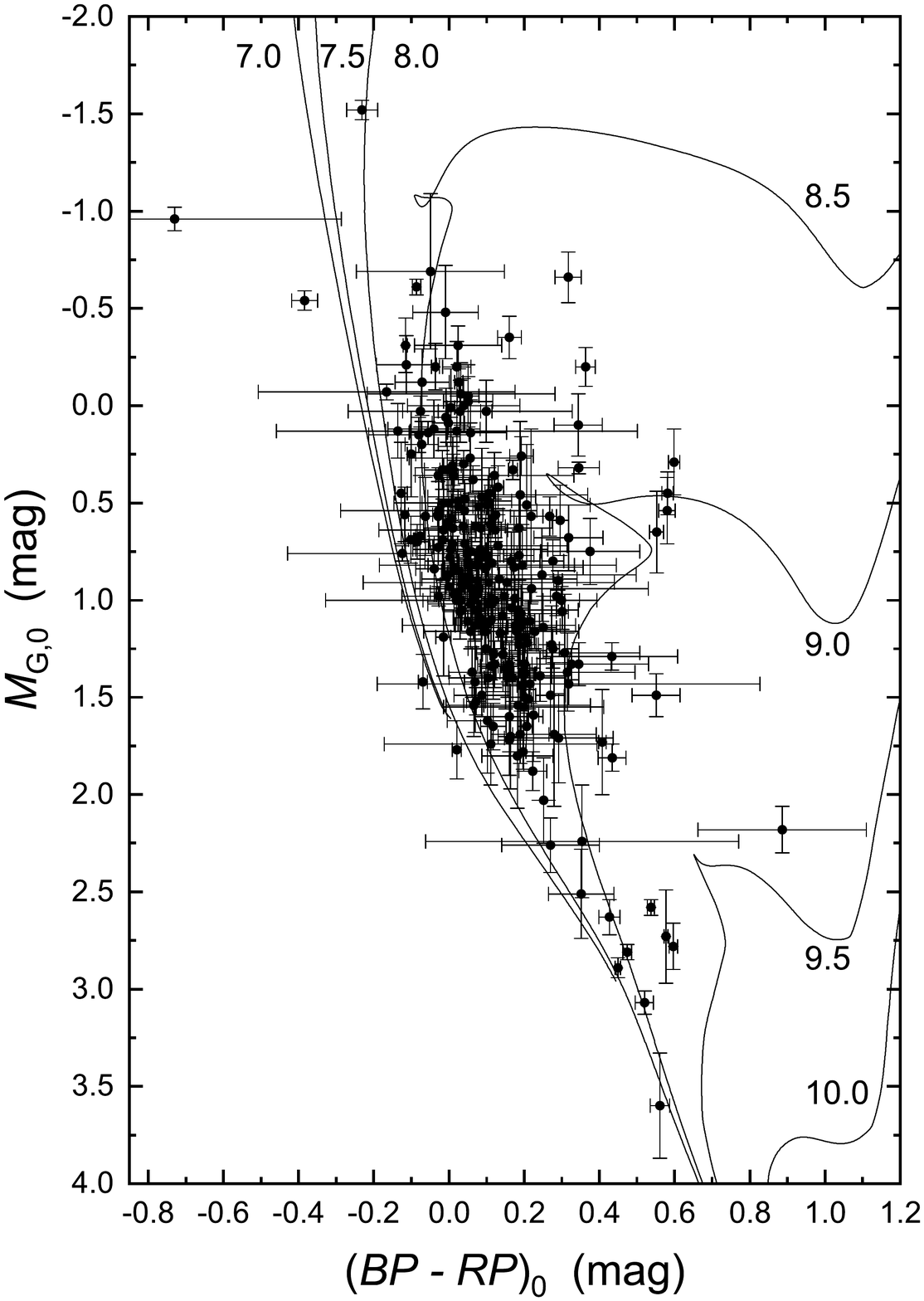}
    \caption{$(BP - RP)_0$ versus $M_{\mathrm{G,0}}$ diagram of our sample stars. Also indicated are PARSEC isochrones for solar metallicity [Z]\,=\,0.02\,dex. Ages are given in logarithmic units of years.}
    \label{fig:CMD}
\end{figure}

The CMD (Figure \ref{fig:CMD}) shows that our sample contains no young stars near the zero-age main sequence (ZAMS). This allowed a corresponding restriction of the tracks in the models to increase the accuracy of the estimates. Parameters could not be determined for five stars of our sample (ATO J076.1702+08.2581, ATO J115.0928-10.6198, ATO J117.8237-09.5101, ATO J308.7048+45.8077, and ATO J321.3283+53.8969) because they did not fit into the grid of models.

\begin{figure}
	\includegraphics[width=\columnwidth,bb=0 0 550 400]{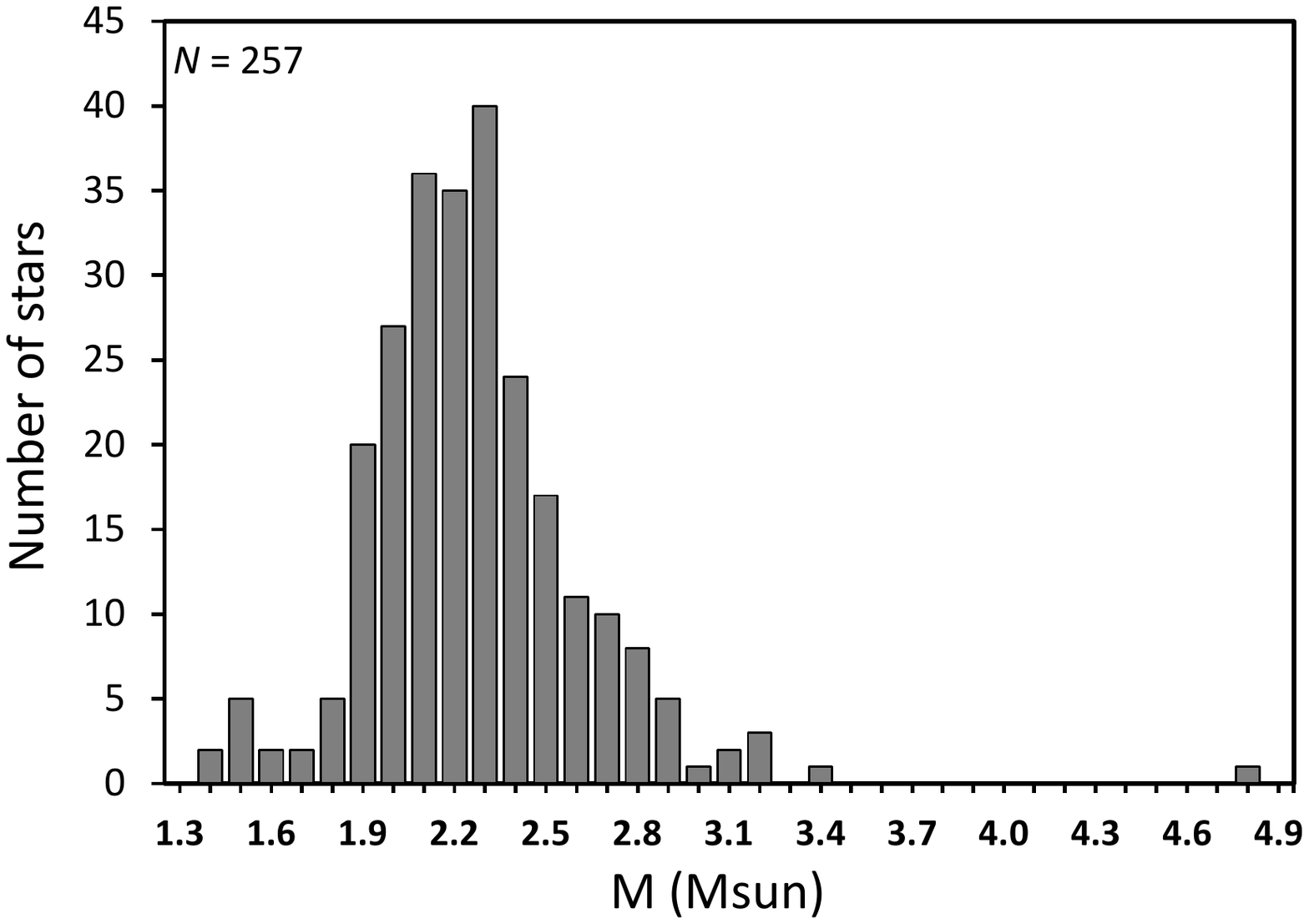}
	\includegraphics[width=\columnwidth,bb=0 0 550 400]{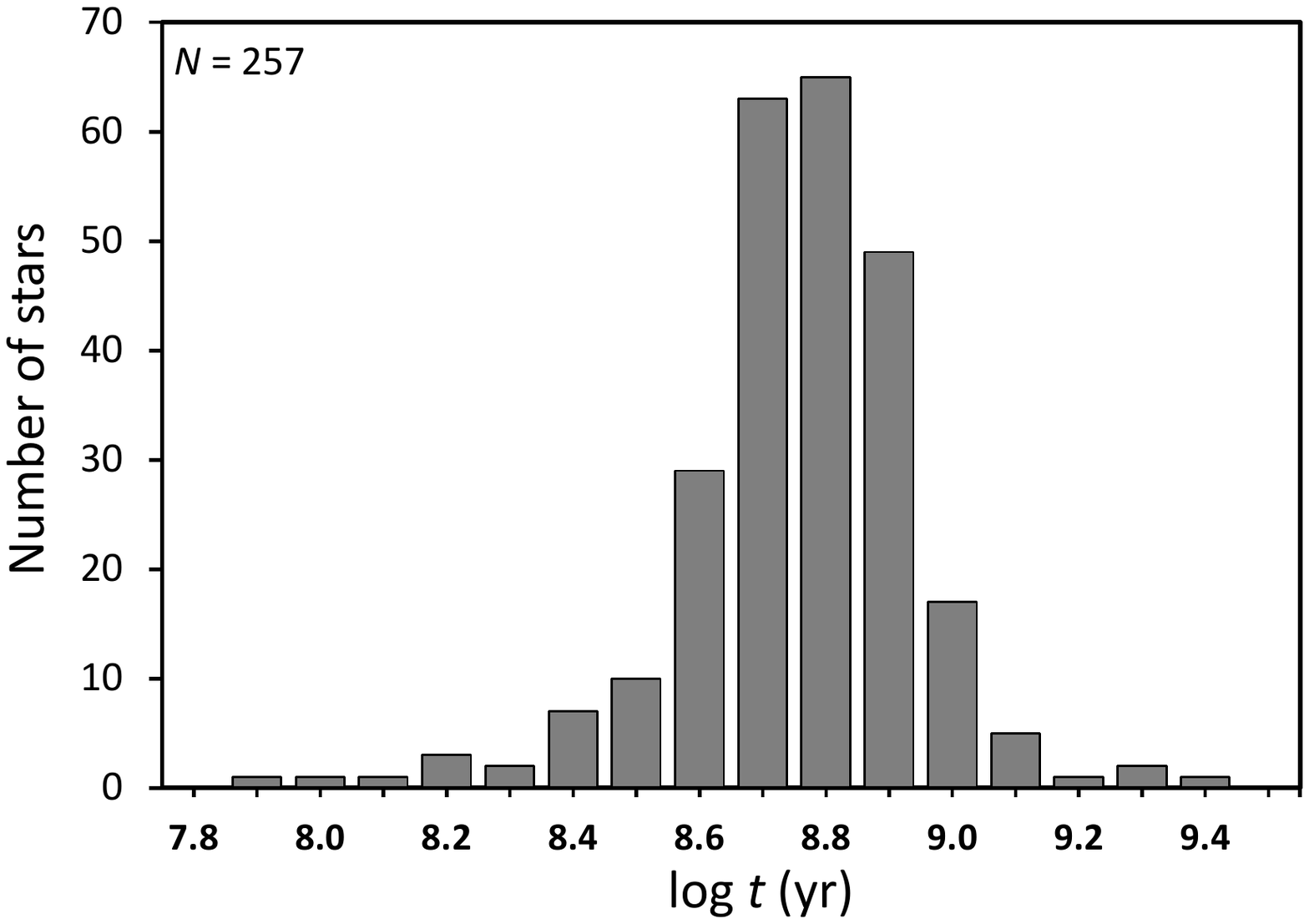}
    \caption{Histograms of the estimated masses (upper panel) and the logarithmic ages (lower panel) of our sample stars.}
    \label{histograms2}
\end{figure}

Figure~\ref{histograms2} shows the mass and age distribution of our sample stars, the complete list of parameters is given in Appendix~\ref{essentialdata}. Derived masses range from 1.4\,\(\textup{M}_\odot\) (a spectral type of about F3) to 5\,\(\textup{M}_\odot\) (B5). However, half of our sample stars have masses between 2 and 2.4\,\(\textup{M}_\odot\) (A2 to B9), which is in line with the spectral classifications presented in Section \ref{spectral_classification}. There is only a single outlier with a mass of 4.8\,\(\textup{M}_\odot\). The derived ages are in agreement with the results from the CMD; the vast majority of our sample stars are main-sequence objects situated rather close to the terminal-age main sequence (TAMS).

\section{Conclusions} \label{conclusions}

Instigated by the similarity of the light curves, periods and amplitudes of the 'upside-down CBH variables' shown in \citet{heinze18} to the parameters of the class of ACV variables, we carried out a search for ACV variables in the ATLAS DR1 catalogue. Our main results are summarised in the following.

\begin{itemize}
	\item Using suitable selection criteria (0.5\,<\,$P(d)$\,<\,10; amp($c$)\,$<$\,0.3\,mag; phase offset $\phi{_2}$\,$-$\,$\phi{_1}$ of the first two Fourier terms between 60$^{\circ}$ and 110$^{\circ}$; stable light curves; monoperiodicity; suitable near-/mid-infrared colours), we identified 264 candidate ACV variables.
	\item 62 of these objects could be spectroscopically confirmed as mCP stars with LAMOST spectra and were classified on the MK system. Nine of these objects are included in the sample of \citet{huemmerich20}; the other 54 stars are new discoveries. The remaining 202 stars are here presented as ACV star candidates that require further observations.
	\item We present basic stellar parameters for all sample stars. Derived masses range from 1.4\,\(\textup{M}_\odot\) to 5\,\(\textup{M}_\odot\), with half our sample stars being situated in the range from 2 \,\(\textup{M}_\odot\) to 2.4 \,\(\textup{M}_\odot\). These results are in good agreement with our spectral classifications. As expected, the majority of our sample stars are main-sequence objects.
	\item Most of our sample stars are members of the thin and thick disk. Four objects (ATO J265.2024+10.6314, ATO J072.9129-07.0638, ATO J005.5419+15.5369, and ATO J310.1614-15.1051) classify as members of the halo population.
	\item With a peak magnitude distribution at around 14th magnitude, the here presented mCP stars and candidates are situated at the faint end of the known Galactic mCP star population.
\end{itemize}

While ACV variables have been included into automated classification routines with good success in the past (e.g. \citealt{sitek14}), several of the more recent, massive variability catalogues, such as \citet{heinze18} and \citet{ZTFvars}, do not specifically consider this class of variable stars. Nevertheless, ACV variables are characterised by a distinct set of parameters in period, amplitude and colour spaces as well as concerning their light curve morphology (monoperiodicity, light curve stability, characteristic phase offset), and should therefore be well suited for automated classification attempts.

Our study shows that a dedicated search for ACV variables in catalogues that do not consider this class is a good starting point for the identification of further members of this group of variables. In general, we expect that the advent of high-precision, multi-colour and spectroscopic surveys will open up a multitude of possibilities to identify new ACV variables, and hence increase the sample size of know mCP stars, in the future.

\section*{Acknowledgements}

We thank the referee for his/her comments that helped to improve the paper. EP acknowledges support by the Erasmus+ programme of the European Union under grant number 2020-1-CZ01-KA203-078200. The Guo Shou Jing Telescope (the Large Sky Area Multi-Object Fiber Spectroscopic Telescope, LAMOST) is a National Major Scientific Project built by the Chinese Academy of Sciences. Funding for the project has been provided by the National Development and Reform Commission. LAMOST is operated and managed by National Astronomical Observatories, Chinese Academy of Sciences. This work presents results from the European Space Agency (ESA) space mission Gaia. Gaia data are being processed by the Gaia Data Processing and Analysis Consortium (DPAC). Funding for the DPAC is provided by national institutions, in particular the institutions participating in the Gaia MultiLateral Agreement (MLA). The Gaia mission website is https://www.cosmos.esa.int/gaia. The Gaia archive website is https://archives.esac.esa.int/gaia.

\section*{Data Availability}
 
The data underlying this article will be shared on reasonable request to the corresponding author.



\bibliographystyle{mnras}
\bibliography{ATLAS} 



\newpage
\appendix

\section{Essential data of our sample stars}\label{essentialdata}

\setcounter{table}{0}
\begin{table*}
\caption{Essential data for our sample stars, sorted by increasing right ascension. The columns denote: (1) ATLAS object identifier. (2) Gaia EDR3 identifier. (3) Right ascension (J2000; Gaia EDR3). (4) Declination (J2000; Gaia EDR3). (5) X coordinate towards the Galactic centre. (6) Y coordinate in direction of Galactic rotation. (7) Z coordinate towards the north Galactic pole. (8) Magnitude in the $G$ band (Gaia EDR3). (9) $G$\,mag error. (10) Rotational period (ATLAS). (11) Median of the photogeometric distance posterior, $D$ (Gaia EDR3). (12) Distance error. (13) Dereddened colour index ($BP-RP$)${_0}$ (Gaia EDR3). (14) Colour index error. (15) Absorption in the $G$ band, $A_G$. (16) Intrinsic absolute magnitude in the $G$ band, M${_G}{_0}$. (17) Absolute magnitude error. (18) Mass. (19) Logarithmic age.}
\label{table_master_part1}
\begin{adjustbox}{max width=\textwidth,angle=90}
\begin{tabular}{lllllllllllllllllll}
\hline
\hline
(1) & (2) & (3) & (4) & (5) & (6) & (7) & (8) & (9) & (10) & (11) & (12) & (13) & (14) & (15) & (16) & (17) & (18) & (19) \\
\hline
ID\_ATO	&	ID\_EDR3	&	RA(J2000)	&	Dec(J2000)	&	X	&	Y	&	Z	&	$G$	&	e\_$G$	&	$P$\_rot	&	$D$	&	e\_$D$	&	($BP-RP$)${_0}$	&	e\_($BP-RP$)${_0}$	&	$A_G$	&	M${_G}{_0}$	&	e\_M${_G}{_0}$	&	Mass	&	$\log t$ \\
	&		&	(hh mm ss.sss)	&	(dd mm ss.sss)	&	(kpc)	&	(kpc)	&	(kpc)	&	(mag)	&	(mag)	&	(d)	&	(pc)	&	(pc)	&	(mag)	&	(mag)	&	(mag)	&	(mag)	&	(mag)	&	(Msun)	&	(yrs) \\
\hline
\hline
J005.5419+15.5369	&	Gaia EDR3 2792368559383811456	&	00 22 10.059	&	+15 32 13.056	&	$-$1.000	&	+2.401	&	$-$2.765	&	13.5507	&	0.0039	&	0.751927	&	3796	&	198	&	+0.581	&	0.015	&	0.21	&	+0.45	&	0.11	&	2.16	&	9.00	\\
J010.7230+57.8087	&	Gaia EDR3 424732034625642368	&	00 42 53.529	&	+57 48 31.384	&	$-$0.924	&	+1.491	&	$-$0.155	&	12.8048	&	0.0038	&	4.435267	&	1761	&	39	&	$-$0.003	&	0.016	&	0.72	&	+0.87	&	0.05	&	2.35	&	8.61	\\
J013.3939+57.6244	&	Gaia EDR3 424250379812359936	&	00 53 34.559	&	+57 37 28.060	&	$-$2.430	&	+3.711	&	$-$0.407	&	13.9679	&	0.0030	&	1.812188	&	4455	&	292	&	$-$0.116	&	0.006	&	1.02	&	$-$0.31	&	0.14	&	3.19	&	8.39	\\
J022.5446+57.3828	&	Gaia EDR3 412876791016632960	&	01 30 10.719	&	+57 22 58.282	&	$-$1.303	&	+1.659	&	$-$0.188	&	14.5231	&	0.0029	&	2.359033	&	2118	&	85	&	+0.197	&	0.007	&	1.10	&	+1.78	&	0.09	&	1.87	&	8.78	\\
J026.6800+56.4974	&	Gaia EDR3 505603588749400192	&	01 46 43.217	&	+56 29 50.893	&	$-$2.698	&	+3.154	&	$-$0.404	&	14.7711	&	0.0033	&	2.146355	&	4170	&	372	&	+0.010	&	0.027	&	1.31	&	+0.36	&	0.20	&	2.52	&	8.67	\\
J031.0380+59.2116	&	Gaia EDR3 507049377823095680	&	02 04 09.142	&	+59 12 41.903	&	$-$1.411	&	+1.560	&	$-$0.086	&	13.7826	&	0.0030	&	2.206660	&	2105	&	59	&	+0.180	&	0.028	&	1.00	&	+1.16	&	0.06	&	2.07	&	8.87	\\
J032.9773+58.5860	&	Gaia EDR3 506819511170582656	&	02 11 54.559	&	+58 35 09.744	&	$-$1.281	&	+1.361	&	$-$0.087	&	12.8886	&	0.0031	&	4.381771	&	1871	&	47	&	+0.101	&	0.011	&	1.02	&	+0.51	&	0.05	&	2.36	&	8.78	\\
J034.4119+58.6875	&	Gaia EDR3 458797241317270912	&	02 17 38.866	&	+58 41 15.122	&	$-$1.786	&	+1.853	&	$-$0.105	&	14.1734	&	0.0031	&	2.475819	&	2575	&	120	&	+0.115	&	0.017	&	1.53	&	+0.58	&	0.10	&	2.32	&	8.80	\\
J038.7653+59.5850	&	Gaia EDR3 465050301379696000	&	02 35 03.677	&	+59 35 06.078	&	$-$1.345	&	+1.312	&	$-$0.023	&	14.0938	&	0.0030	&	1.507414	&	1880	&	50	&	+0.054	&	0.023	&	1.82	&	+0.88	&	0.06	&	2.26	&	8.73	\\
J039.8159+57.4010	&	Gaia EDR3 457901276777494016	&	02 39 15.825	&	+57 24 03.870	&	$-$1.986	&	+1.847	&	$-$0.117	&	15.0181	&	0.0032	&	1.129681	&	2714	&	162	&	+0.195	&	0.013	&	1.47	&	+1.37	&	0.13	&	1.99	&	8.88	\\
J041.5233+57.1487	&	Gaia EDR3 460845975073772416	&	02 46 05.591	&	+57 08 55.459	&	$-$2.244	&	+2.019	&	$-$0.123	&	15.5073	&	0.0031	&	1.568993	&	3021	&	254	&	+0.296	&	0.022	&	2.52	&	+0.59	&	0.18	&	2.25	&	8.89	\\
J042.3277+57.1063	&	Gaia EDR3 460790655894929280	&	02 49 18.660	&	+57 06 22.823	&	$-$1.979	&	+1.755	&	$-$0.100	&	15.0568	&	0.0030	&	1.697997	&	2647	&	176	&	+0.062	&	0.023	&	1.99	&	+0.94	&	0.15	&	2.23	&	8.74	\\
J042.3381+51.3632	&	Gaia EDR3 440615407799027840	&	02 49 21.164	&	+51 21 47.553	&	$-$1.798	&	+1.457	&	$-$0.298	&	13.5526	&	0.0039	&	2.762592	&	2333	&	97	&	+0.042	&	0.012	&	0.83	&	+0.88	&	0.09	&	2.27	&	8.71	\\
J046.2665+57.4649	&	Gaia EDR3 460271445887343104	&	03 05 03.984	&	+57 27 53.658	&	$-$1.765	&	+1.473	&	$-$0.035	&	13.9203	&	0.0031	&	2.959672	&	2299	&	86	&	$-$0.076	&	0.047	&	2.06	&	+0.03	&	0.08	&	2.87	&	8.50	\\
J046.9585+53.9453	&	Gaia EDR3 447101529969139200	&	03 07 50.055	&	+53 56 43.237	&	$-$1.621	&	+1.255	&	$-$0.134	&	12.7750	&	0.0029	&	1.610395	&	2054	&	103	&	+0.161	&	0.032	&	1.55	&	$-$0.35	&	0.11	&	2.66	&	8.74	\\
J047.5021+53.3159	&	Gaia EDR3 446269337107134720	&	03 10 00.500	&	+53 18 57.326	&		&		&		&	13.4847	&	0.0095	&	2.289972	&		&		&		&		&		&		&		&		&		\\
J050.9303+44.3793	&	Gaia EDR3 241472445889902080	&	03 23 43.271	&	+44 22 45.440	&	$-$3.261	&	+1.917	&	$-$0.700	&	14.6146	&	0.0046	&	4.494699	&	3847	&	362	&	$-$0.015	&	0.020	&	0.46	&	+1.19	&	0.20	&	2.27	&	8.41	\\
J053.2447+44.4347	&	Gaia EDR3 241650223177124864	&	03 32 58.738	&	+44 26 05.071	&	$-$1.199	&	+0.667	&	$-$0.229	&	12.1581	&	0.0031	&	1.472468	&	1392	&	36	&	+0.102	&	0.009	&	0.42	&	+1.02	&	0.06	&	2.16	&	8.79	\\
J057.4910+58.7724	&	Gaia EDR3 473203695496163200	&	03 49 57.859	&	+58 46 20.852	&	$-$1.373	&	+0.985	&	+0.104	&	13.4440	&	0.0029	&	2.675655	&	1693	&	31	&	+0.084	&	0.015	&	1.51	&	+0.78	&	0.04	&	2.27	&	8.77	\\
J057.6555+58.8757	&	Gaia EDR3 473205963238846848	&	03 50 37.322	&	+58 52 32.644	&	$-$2.695	&	+1.934	&	+0.212	&	15.7241	&	0.0035	&	1.803022	&	3324	&	373	&	+0.225	&	0.014	&	1.50	&	+1.59	&	0.24	&	1.90	&	8.89	\\
J057.9558+54.6451	&	Gaia EDR3 444608868392387456	&	03 51 49.391	&	+54 38 42.453	&	$-$0.974	&	+0.629	&	+0.010	&	13.2747	&	0.0031	&	2.222517	&	1160	&	59	&	+0.345	&	0.186	&	1.55	&	+1.33	&	0.11	&	1.94	&	9.01	\\
J059.0278+55.2600	&	Gaia EDR3 444653879648018816	&	03 56 06.680	&	+55 15 36.286	&	$-$1.152	&	+0.742	&	+0.032	&	12.6441	&	0.0028	&	1.824059	&	1371	&	23	&	+0.123	&	0.016	&	1.31	&	+0.64	&	0.04	&	2.29	&	8.81	\\
J059.8990+50.2781	&	Gaia EDR3 250585816735404544	&	03 59 35.769	&	+50 16 41.264	&	$-$2.628	&	+1.467	&	$-$0.111	&	15.7401	&	0.0044	&	4.764477	&	3012	&	272	&	+0.273	&	0.035	&	2.10	&	+1.23	&	0.20	&	2.00	&	8.95	\\
J060.4363+55.5067	&	Gaia EDR3 468500667647878400	&	04 01 44.730	&	+55 30 24.406	&	$-$1.492	&	+0.945	&	+0.063	&	13.7076	&	0.0034	&	2.474011	&	1767	&	47	&	+0.176	&	0.020	&	1.46	&	+0.99	&	0.06	&	2.13	&	8.87	\\
J060.7275+46.2448	&	Gaia EDR3 245658458458317056	&	04 02 54.602	&	+46 14 41.536	&	$-$3.839	&	+1.880	&	$-$0.358	&	15.7085	&	0.0036	&	1.240184	&	4289	&	555	&	+0.098	&	0.047	&	1.27	&	+1.25	&	0.28	&	2.09	&	8.76	\\
J060.7480+45.3172	&	Gaia EDR3 233468173039182464	&	04 02 59.532	&	+45 19 02.115	&	$-$1.368	&	+0.651	&	$-$0.145	&	13.0578	&	0.0029	&	1.522352	&	1522	&	58	&	+0.137	&	0.006	&	0.98	&	+1.17	&	0.08	&	2.09	&	8.83	\\
J061.6000+59.6651	&	Gaia EDR3 470578302247842560	&	04 06 24.001	&	+59 39 54.612	&	$-$2.623	&	+1.813	&	+0.311	&	14.0773	&	0.0042	&	3.434188	&	3204	&	201	&	$-$0.005	&	0.019	&	1.21	&	+0.33	&	0.14	&	2.56	&	8.65	\\
J062.2757+57.3439	&	Gaia EDR3 469300326139460224	&	04 09 06.182	&	+57 20 38.089	&	$-$2.820	&	+1.818	&	+0.240	&	14.5011	&	0.0034	&	2.119598	&	3364	&	195	&	+0.107	&	0.011	&	1.41	&	+0.45	&	0.13	&	2.38	&	8.78	\\
J062.6054+48.7741	&	Gaia EDR3 246505838323965952	&	04 10 25.302	&	+48 46 26.922	&	$-$2.601	&	+1.317	&	$-$0.105	&	15.7129	&	0.0034	&	3.863680	&	2917	&	345	&	+0.086	&	0.210	&	2.23	&	+1.13	&	0.26	&	2.14	&	8.76	\\
J063.5805+46.9075	&	Gaia EDR3 234065826327540608	&	04 14 19.331	&	+46 54 27.154	&	$-$2.368	&	+1.109	&	$-$0.136	&	13.1824	&	0.0037	&	1.456892	&	2618	&	499	&	$-$0.049	&	0.197	&	1.72	&	$-$0.69	&	0.40	&	3.17	&	8.49	\\
J064.2611+46.2848	&	Gaia EDR3 233792632048827776	&	04 17 02.662	&	+46 17 05.502	&	$-$2.461	&	+1.112	&	$-$0.146	&	14.7410	&	0.0032	&	2.681701	&	2704	&	234	&	+0.119	&	0.030	&	1.26	&	+1.28	&	0.19	&	2.06	&	8.79	\\
J065.5257+51.2992	&	Gaia EDR3 271532063001936512	&	04 22 06.167	&	+51 17 57.385	&	$-$2.448	&	+1.261	&	+0.051	&	14.8359	&	0.0046	&	3.622927	&	2754	&	197	&	+0.005	&	0.094	&	1.85	&	+0.74	&	0.15	&	2.38	&	8.66	\\
J065.7038+47.6938	&	Gaia EDR3 257945019859374208	&	04 22 48.930	&	+47 41 38.038	&	$-$0.712	&	+0.326	&	$-$0.019	&	11.4438	&	0.0038	&	2.774777	&	783	&	109	&	+0.154	&	0.021	&	1.02	&	+0.91	&	0.30	&	2.17	&	8.85	\\
J065.8718+43.5268	&	Gaia EDR3 229348646566881664	&	04 23 29.244	&	+43 31 36.752	&	$-$2.989	&	+1.181	&	$-$0.239	&	14.3782	&	0.0035	&	1.453359	&	3223	&	193	&	+0.058	&	0.044	&	1.09	&	+0.75	&	0.13	&	2.30	&	8.74	\\
J066.8944+56.8176	&	Gaia EDR3 278208606841140864	&	04 27 34.661	&	+56 49 03.467	&	$-$3.167	&	+1.876	&	+0.353	&	14.2141	&	0.0031	&	1.708314	&	3698	&	233	&	$-$0.136	&	0.032	&	1.22	&	+0.13	&	0.14	&	3.08	&	8.27	\\
J066.9854+44.3739	&	Gaia EDR3 253427023504582016	&	04 27 56.516	&	+44 22 26.160	&	$-$2.820	&	+1.117	&	$-$0.164	&	14.3520	&	0.0029	&	1.716098	&	3038	&	182	&	+0.186	&	0.040	&	1.14	&	+0.77	&	0.13	&	2.21	&	8.86	\\
J067.0419+51.6124	&	Gaia EDR3 271176817665195008	&	04 28 10.072	&	+51 36 44.750	&	$-$3.565	&	+1.802	&	+0.136	&	15.5832	&	0.0042	&	6.716061	&	3997	&	545	&	+0.173	&	0.184	&	1.74	&	+0.83	&	0.29	&	2.19	&	8.86	\\
J067.4887+38.9535	&	Gaia EDR3 179588117545396992	&	04 29 57.299	&	+38 57 12.665	&	$-$2.593	&	+0.813	&	$-$0.312	&	14.3917	&	0.0034	&	3.009545	&	2735	&	198	&	+0.038	&	0.009	&	1.59	&	+0.62	&	0.16	&	2.37	&	8.72	\\
J067.6640+38.2381	&	Gaia EDR3 178430469239195904	&	04 30 39.368	&	+38 14 17.516	&	$-$2.235	&	+0.674	&	$-$0.284	&	13.7970	&	0.0032	&	1.387788	&	2352	&	100	&	+0.042	&	0.019	&	1.23	&	+0.71	&	0.09	&	2.34	&	8.72	\\
J072.5642+39.5294	&	Gaia EDR3 199942929634159360	&	04 50 15.413	&	+39 31 46.032	&	$-$3.716	&	+1.012	&	$-$0.216	&	14.6053	&	0.0036	&	2.263109	&	3857	&	799	&	+0.219	&	0.055	&	1.04	&	+0.57	&	0.45	&	2.28	&	8.86	\\
J072.9129-07.0638	&	Gaia EDR3 3187615948455925120	&	04 51 39.098	&	-07 03 49.730	&	$-$3.411	&	$-$1.606	&	$-$2.159	&	13.9441	&	0.0048	&	0.734840	&	4345	&	343	&	+0.580	&	0.022	&	0.20	&	+0.54	&	0.17	&	2.12	&	9.02	\\
J073.7460+43.3008	&	Gaia EDR3 205240445379721728	&	04 54 59.044	&	+43 18 02.990	&	$-$2.590	&	+0.821	&	$-$0.007	&	13.5702	&	0.0039	&	7.746157	&	2717	&	125	&	+0.049	&	0.013	&	0.65	&	+0.76	&	0.10	&	2.31	&	8.73	\\
J073.8114+47.4553	&	Gaia EDR3 254938577111634560	&	04 55 14.756	&	+47 27 19.237	&	$-$2.679	&	+1.017	&	+0.125	&	13.7983	&	0.0029	&	2.811480	&	2868	&	127	&	+0.009	&	0.024	&	1.19	&	+0.31	&	0.10	&	2.54	&	8.67	\\
J074.4268+43.8711	&	Gaia EDR3 205351083735634688	&	04 57 42.450	&	+43 52 16.107	&	$-$2.179	&	+0.696	&	+0.024	&	13.8735	&	0.0030	&	4.774356	&	2287	&	101	&	+0.286	&	0.016	&	1.09	&	+0.98	&	0.10	&	2.09	&	8.94	\\
J074.6123+26.0721	&	Gaia EDR3 153687849841184768	&	04 58 26.967	&	+26 04 19.737	&	$-$2.324	&	+0.144	&	$-$0.421	&	14.4723	&	0.0030	&	2.610292	&	2366	&	112	&	$-$0.003	&	0.006	&	1.98	&	+0.60	&	0.10	&	2.44	&	8.65	\\
J074.6661+44.7655	&	Gaia EDR3 205467803765798912	&	04 58 39.883	&	+44 45 55.797	&	$-$2.002	&	+0.663	&	+0.048	&	14.1854	&	0.0036	&	1.675826	&	2109	&	78	&	+0.203	&	0.017	&	1.36	&	+1.19	&	0.08	&	2.05	&	8.89	\\
J074.9570+40.2063	&	Gaia EDR3 200775706615199616	&	04 59 49.685	&	+40 12 22.783	&	$-$3.100	&	+0.808	&	$-$0.076	&	14.7759	&	0.0037	&	1.557928	&	3204	&	234	&	+0.112	&	0.067	&	1.14	&	+1.10	&	0.16	&	2.13	&	8.80	\\
J074.9889+42.2979	&	Gaia EDR3 201974204357706496	&	04 59 57.342	&	+42 17 52.556	&	$-$3.775	&	+1.099	&	$-$0.003	&	15.5504	&	0.0033	&	3.659410	&	3932	&	492	&	+0.102	&	0.107	&	0.91	&	+1.62	&	0.27	&	1.98	&	8.60	\\
J075.0682+42.3262	&	Gaia EDR3 201974483527413248	&	05 00 16.368	&	+42 19 34.383	&	$-$2.453	&	+0.714	&	+0.001	&	14.4160	&	0.0035	&	1.863155	&	2554	&	140	&	+0.118	&	0.034	&	0.72	&	+1.65	&	0.12	&	1.96	&	8.64	\\
J075.5525+46.9691	&	Gaia EDR3 207004336905968896	&	05 02 12.602	&	+46 58 09.022	&	$-$2.283	&	+0.816	&	+0.133	&	14.4067	&	0.0033	&	2.664824	&	2428	&	149	&	+0.104	&	0.063	&	1.96	&	+0.49	&	0.13	&	2.37	&	8.78	\\
J075.7839+42.1896	&	Gaia EDR3 201793957461699200	&	05 03 08.148	&	+42 11 22.744	&	$-$3.917	&	+1.107	&	+0.025	&	13.5006	&	0.0030	&	1.458149	&	4070	&	180	&	+0.363	&	0.026	&	0.64	&	$-$0.20	&	0.10	&	2.49	&	8.83	\\
J076.1702+08.2581	&	Gaia EDR3 3290096582558387072	&	05 04 40.867	&	+08 15 29.218	&	$-$2.438	&	$-$0.538	&	$-$0.873	&	16.0511	&	0.0049	&	2.318806	&	2645	&	341	&	+0.561	&	0.026	&	0.29	&	+3.60	&	0.27	&		&		\\
J076.3581+43.8370	&	Gaia EDR3 202507403068125312	&	05 05 25.948	&	+43 50 13.197	&	$-$2.773	&	+0.839	&	+0.085	&	15.1060	&	0.0031	&	2.363328	&	2898	&	231	&	+0.375	&	0.132	&	2.02	&	+0.75	&	0.17	&	2.16	&	8.95	\\
J076.4023+45.6101	&	Gaia EDR3 205975262742313984	&	05 05 36.568	&	+45 36 36.600	&	$-$2.587	&	+0.852	&	+0.132	&	14.5110	&	0.0039	&	2.156734	&	2727	&	161	&	+0.078	&	0.036	&	1.28	&	+1.05	&	0.13	&	2.17	&	8.75	\\
J076.5930+30.8311	&	Gaia EDR3 157208722295230208	&	05 06 22.337	&	+30 49 52.004	&	$-$2.412	&	+0.268	&	$-$0.255	&	14.0221	&	0.0033	&	1.514704	&	2440	&	124	&	+0.142	&	0.011	&	0.99	&	+1.08	&	0.11	&	2.11	&	8.84	\\
J077.0867+36.5911	&	Gaia EDR3 185732154163148160	&	05 08 20.818	&	+36 35 28.032	&	$-$2.863	&	+0.542	&	$-$0.113	&	14.4993	&	0.0031	&	1.929063	&	2916	&	217	&	+0.072	&	0.082	&	1.17	&	+0.96	&	0.16	&	2.21	&	8.75	\\
J077.1846+45.6644	&	Gaia EDR3 205952340498929664	&	05 08 44.317	&	+45 39 52.175	&	$-$2.652	&	+0.859	&	+0.158	&	13.4514	&	0.0031	&	4.835060	&	2792	&	107	&	+0.079	&	0.015	&	0.70	&	+0.52	&	0.08	&	2.37	&	8.76	\\
J077.7414+40.6851	&	Gaia EDR3 201008871800476288	&	05 10 57.948	&	+40 41 06.642	&	$-$3.948	&	+0.963	&	+0.044	&	13.7868	&	0.0029	&	1.678861	&	4064	&	319	&	$-$0.071	&	0.072	&	0.84	&	$-$0.12	&	0.17	&	2.92	&	8.51	\\
J079.1356+33.7630	&	Gaia EDR3 181901623151346304	&	05 16 32.551	&	+33 45 46.825	&	$-$2.699	&	+0.353	&	$-$0.121	&	15.9835	&	0.0041	&	2.458936	&	2725	&	289	&	+0.352	&	0.087	&	1.28	&	+2.51	&	0.23	&	1.62	&	8.64	\\
J079.9036+53.6898	&	Gaia EDR3 266385317788362240	&	05 19 36.880	&	+53 41 23.462	&	$-$3.210	&	+1.398	&	+0.574	&	14.9573	&	0.0034	&	2.926359	&	3548	&	343	&	+0.151	&	0.005	&	0.87	&	+1.35	&	0.21	&	2.02	&	8.82	\\
J080.3836+43.4165	&	Gaia EDR3 207392086553729408	&	05 21 32.067	&	+43 24 59.551	&	$-$4.424	&	+1.168	&	+0.304	&	14.4180	&	0.0032	&	2.896787	&	4586	&	418	&	+0.005	&	0.010	&	1.08	&	+0.01	&	0.20	&	2.69	&	8.65	\\
J080.5363+36.3112	&	Gaia EDR3 183986301496148096	&	05 22 08.724	&	+36 18 40.684	&	$-$2.718	&	+0.424	&	$-$0.007	&	15.5640	&	0.0045	&	1.679479	&	2751	&	327	&	+0.184	&	0.192	&	1.82	&	+1.54	&	0.26	&	1.94	&	8.84	\\
J081.6900+35.0200	&	Gaia EDR3 182878543531780480	&	05 26 45.607	&	+35 01 12.138	&	$-$1.801	&	+0.230	&	$-$0.003	&	13.4568	&	0.0029	&	0.804327	&	1815	&	66	&	+0.025	&	0.081	&	1.61	&	+0.52	&	0.08	&	2.43	&	8.70	\\
J081.9629+42.4325	&	Gaia EDR3 195375450956448768	&	05 27 51.121	&	+42 25 57.350	&	$-$1.687	&	+0.400	&	+0.128	&	12.9573	&	0.0050	&	1.761931	&	1739	&	58	&	+0.070	&	0.021	&	0.62	&	+1.13	&	0.07	&	2.16	&	8.73	\\
J082.4146+43.5804	&	Gaia EDR3 195691564846752128	&	05 29 39.510	&	+43 34 49.729	&	$-$3.558	&	+0.894	&	+0.328	&	15.3747	&	0.0036	&	2.454385	&	3684	&	401	&	+0.292	&	0.144	&	0.82	&	+1.71	&	0.23	&	1.84	&	8.96	\\
J082.4358+39.0290	&	Gaia EDR3 190674974325136384	&	05 29 44.596	&	+39 01 44.550	&	$-$2.514	&	+0.456	&	+0.117	&	14.7324	&	0.0046	&	1.947387	&	2558	&	153	&	+0.242	&	0.074	&	1.28	&	+1.39	&	0.13	&	1.96	&	8.93	\\
J082.5579+33.1925	&	Gaia EDR3 3449031263234277376	&	05 30 13.910	&	+33 11 33.315	&	$-$2.477	&	+0.232	&	$-$0.022	&	13.4316	&	0.0033	&	4.734125	&	2488	&	91	&	+0.011	&	0.076	&	1.11	&	+0.33	&	0.08	&	2.53	&	8.68	\\
J082.6115+29.6992	&	Gaia EDR3 3446133637776241792	&	05 30 26.782	&	+29 41 57.238	&	$-$2.679	&	+0.113	&	$-$0.112	&	14.6350	&	0.0036	&	1.963054	&	2684	&	178	&	+0.105	&	0.071	&	1.34	&	+1.12	&	0.14	&	2.13	&	8.79	\\
J082.6322+21.1693	&	Gaia EDR3 3403667578134303616	&	05 30 31.744	&	+21 10 09.472	&	$-$4.845	&	$-$0.404	&	$-$0.599	&	14.4688	&	0.0033	&	2.215640	&	4899	&	498	&	$-$0.100	&	0.011	&	0.77	&	+0.25	&	0.22	&	2.85	&	8.41	\\
J083.0553+11.5461	&	Gaia EDR3 3340653192355972480	&	05 32 13.292	&	+11 32 46.117	&	$-$1.144	&	$-$0.270	&	$-$0.245	&	14.0975	&	0.0031	&	2.221372	&	1200	&	35	&	+0.520	&	0.024	&	0.62	&	+3.07	&	0.06	&	1.43	&	8.88	\\
\hline
\end{tabular}                                                                                                                                                                   
\end{adjustbox}
\end{table*}

\setcounter{table}{0}
\begin{table*}
\caption{Essential data for our sample stars, sorted by increasing right ascension. The columns denote: (1) ATLAS object identifier. (2) Gaia EDR3 identifier. (3) Right ascension (J2000; Gaia EDR3). (4) Declination (J2000; Gaia EDR3). (5) X coordinate towards the Galactic centre. (6) Y coordinate in direction of Galactic rotation. (7) Z coordinate towards the north Galactic pole. (8) Magnitude in the $G$ band (Gaia EDR3). (9) $G$\,mag error. (10) Rotational period (ATLAS). (11) Median of the photogeometric distance posterior, $D$ (Gaia EDR3). (12) Distance error. (13) Dereddened colour index ($BP-RP$)${_0}$ (Gaia EDR3). (14) Colour index error. (15) Absorption in the $G$ band, $A_G$. (16) Intrinsic absolute magnitude in the $G$ band, M${_G}{_0}$. (17) Absolute magnitude error. (18) Mass. (19) Logarithmic age.}
\label{table_master_part2}
\begin{adjustbox}{max width=\textwidth,angle=90}
\begin{tabular}{lllllllllllllllllll}
\hline
\hline
(1) & (2) & (3) & (4) & (5) & (6) & (7) & (8) & (9) & (10) & (11) & (12) & (13) & (14) & (15) & (16) & (17) & (18) & (19) \\
\hline
ID\_ATO	&	ID\_EDR3	&	RA(J2000)	&	Dec(J2000)	&	X	&	Y	&	Z	&	$G$	&	e\_$G$	&	$P$\_rot	&	$D$	&	e\_$D$	&	($BP-RP$)${_0}$	&	e\_($BP-RP$)${_0}$	&	$A_G$	&	M${_G}{_0}$	&	e\_M${_G}{_0}$	&	Mass	&	$\log t$ \\
	&		&	(hh mm ss.sss)	&	(dd mm ss.sss)	&	(kpc)	&	(kpc)	&	(kpc)	&	(mag)	&	(mag)	&	(d)	&	(pc)	&	(pc)	&	(mag)	&	(mag)	&	(mag)	&	(mag)	&	(mag)	&	(Msun)	&	(yrs) \\
\hline
\hline
J083.1990+35.5978	&	Gaia EDR3 183188189196465152	&	05 32 47.768	&	+35 35 52.246	&	$-$2.866	&	+0.356	&	+0.063	&	14.4956	&	0.0038	&	1.637762	&	2889	&	178	&	+0.020	&	0.089	&	1.17	&	+1.00	&	0.13	&	2.27	&	8.64	\\
J084.6836+33.5786	&	Gaia EDR3 3449257139854626944	&	05 38 44.070	&	+33 34 43.039	&	$-$2.177	&	+0.180	&	+0.046	&	13.8926	&	0.0033	&	1.083538	&	2185	&	97	&	+0.201	&	0.033	&	0.85	&	+1.33	&	0.10	&	2.00	&	8.89	\\
J084.9321+37.2119	&	Gaia EDR3 189377859842561664	&	05 39 43.724	&	+37 12 43.056	&	$-$2.024	&	+0.273	&	+0.118	&	14.6232	&	0.0057	&	1.838578	&	2046	&	116	&	+0.122	&	0.045	&	1.70	&	+1.33	&	0.12	&	2.05	&	8.78	\\
J086.1147+31.8873	&	Gaia EDR3 3448105680601602048	&	05 44 27.548	&	+31 53 14.317	&	$-$2.807	&	+0.130	&	+0.065	&	15.2939	&	0.0030	&	1.932094	&	2811	&	283	&	+0.252	&	0.013	&	1.00	&	+2.03	&	0.22	&	1.77	&	8.78	\\
J086.1972+24.3919	&	Gaia EDR3 3428577117142074880	&	05 44 47.349	&	+24 23 31.095	&	$-$4.039	&	$-$0.267	&	$-$0.179	&	14.0418	&	0.0029	&	1.818867	&	4052	&	434	&	$-$0.009	&	0.087	&	1.50	&	$-$0.48	&	0.24	&	2.97	&	8.57	\\
J086.3082+11.2506	&	Gaia EDR3 3337017760238097664	&	05 45 13.974	&	+11 15 02.442	&	$-$1.615	&	$-$0.438	&	$-$0.271	&	12.0344	&	0.0030	&	2.122828	&	1695	&	57	&	+0.026	&	0.011	&	1.00	&	$-$0.12	&	0.07	&	2.72	&	8.65	\\
J086.4762+23.3391	&	Gaia EDR3 3427626413242761856	&	05 45 54.299	&	+23 20 21.113	&	$-$1.544	&	$-$0.130	&	$-$0.077	&	13.7338	&	0.0031	&	2.203900	&	1552	&	56	&	+0.326	&	0.105	&	1.43	&	+1.33	&	0.08	&	1.95	&	9.00	\\
J086.5553+40.4840	&	Gaia EDR3 191794242803169920	&	05 46 13.292	&	+40 29 02.749	&	$-$1.886	&	+0.327	&	+0.204	&	13.0681	&	0.0029	&	1.633911	&	1925	&	80	&	+0.123	&	0.007	&	0.62	&	+1.00	&	0.09	&	2.15	&	8.82	\\
J086.8565+27.9110	&	Gaia EDR3 3443080736365795456	&	05 47 25.565	&	+27 54 39.651	&	$-$3.241	&	$-$0.061	&	$-$0.011	&	15.0779	&	0.0042	&	3.586864	&	3242	&	357	&	+0.247	&	0.251	&	1.62	&	+0.87	&	0.24	&	2.14	&	8.91	\\
J087.0695+34.3930	&	Gaia EDR3 3454688662875499520	&	05 48 16.696	&	+34 23 35.073	&	$-$1.727	&	+0.132	&	+0.100	&	14.4142	&	0.0029	&	1.941033	&	1735	&	82	&	+0.223	&	0.037	&	1.32	&	+1.88	&	0.10	&	1.83	&	8.79	\\
J087.1757+29.1833	&	Gaia EDR3 3443457426475467136	&	05 48 42.185	&	+29 11 00.093	&	$-$1.420	&	$-$0.003	&	+0.017	&	14.2574	&	0.0038	&	1.676034	&	1420	&	190	&	+0.354	&	0.416	&	1.00	&	+2.24	&	0.29	&	1.67	&	8.94	\\
J087.2218+29.4548	&	Gaia EDR3 3443533052258430976	&	05 48 53.234	&	+29 27 17.365	&	$-$2.081	&	+0.003	&	+0.032	&	13.7611	&	0.0030	&	1.103880	&	2081	&	83	&	+0.228	&	0.117	&	0.99	&	+1.16	&	0.09	&	2.04	&	8.92	\\
J087.2751+25.3155	&	Gaia EDR3 3428987093240848640	&	05 49 06.044	&	+25 18 55.885	&	$-$1.248	&	$-$0.076	&	$-$0.026	&	11.9445	&	0.0039	&	1.495326	&	1251	&	50	&	+0.027	&	0.047	&	0.59	&	+0.85	&	0.09	&	2.30	&	8.69	\\
J088.2315+38.6528	&	Gaia EDR3 3457964004934635136	&	05 52 55.568	&	+38 39 10.186	&	$-$3.339	&	+0.445	&	+0.370	&	13.6732	&	0.0030	&	1.840969	&	3389	&	238	&	$-$0.008	&	0.013	&	0.97	&	+0.06	&	0.15	&	2.68	&	8.63	\\
J089.0960+24.8012	&	Gaia EDR3 3428042754490731776	&	05 56 23.041	&	+24 48 04.511	&	$-$3.179	&	$-$0.266	&	$-$0.003	&	15.4828	&	0.0032	&	1.363957	&	3190	&	284	&	+0.143	&	0.040	&	1.67	&	+1.28	&	0.19	&	2.05	&	8.82	\\
J089.6385+19.6248	&	Gaia EDR3 3398736298547472768	&	05 58 33.253	&	+19 37 29.648	&	$-$2.518	&	$-$0.422	&	$-$0.099	&	14.4676	&	0.0030	&	2.940504	&	2555	&	152	&	+0.177	&	0.160	&	1.28	&	+1.13	&	0.13	&	2.08	&	8.87	\\
J089.7490+22.3852	&	Gaia EDR3 3424448416618489216	&	05 58 59.776	&	+22 23 06.916	&	$-$2.729	&	$-$0.343	&	$-$0.036	&	14.4788	&	0.0031	&	1.828602	&	2751	&	169	&	+0.063	&	0.089	&	1.26	&	+1.00	&	0.13	&	2.21	&	8.73	\\
J089.7821+14.0001	&	Gaia EDR3 3343881599015125248	&	05 59 07.711	&	+14 00 00.667	&	$-$3.347	&	$-$0.864	&	$-$0.295	&	14.0068	&	0.0044	&	2.914186	&	3469	&	217	&	$-$0.118	&	0.010	&	0.77	&	+0.56	&	0.14	&	2.82	&	8.19	\\
J090.0237+21.9017	&	Gaia EDR3 3423296605171557504	&	06 00 05.708	&	+21 54 06.341	&	$-$2.155	&	$-$0.292	&	$-$0.029	&	14.6374	&	0.0030	&	1.852552	&	2174	&	112	&	+0.160	&	0.046	&	1.60	&	+1.33	&	0.11	&	2.02	&	8.84	\\
J090.3141+14.7489	&	Gaia EDR3 3345435449462667776	&	06 01 15.396	&	+14 44 56.346	&	$-$3.380	&	$-$0.847	&	$-$0.248	&	14.8482	&	0.0031	&	2.366286	&	3493	&	330	&	+0.111	&	0.069	&	1.48	&	+0.61	&	0.20	&	2.31	&	8.80	\\
J090.7455+31.0548	&	Gaia EDR3 3449912762321090176	&	06 02 58.925	&	+31 03 17.267	&	$-$3.970	&	$-$0.005	&	+0.299	&	14.8417	&	0.0031	&	1.918545	&	3981	&	490	&	$-$0.127	&	0.016	&	1.36	&	+0.45	&	0.26	&	2.91	&	8.18	\\
J090.7799+32.5180	&	Gaia EDR3 3450617063939465088	&	06 03 07.176	&	+32 31 05.063	&	$-$3.655	&	+0.076	&	+0.323	&	15.6062	&	0.0046	&	3.392613	&	3670	&	456	&	+0.164	&	0.034	&	1.08	&	+1.70	&	0.27	&	1.91	&	8.74	\\
J091.1746+24.8696	&	Gaia EDR3 3426554634224242816	&	06 04 41.925	&	+24 52 10.875	&	$-$4.213	&	$-$0.417	&	+0.120	&	15.7398	&	0.0037	&	2.587859	&	4235	&	645	&	+0.199	&	0.213	&	1.05	&	+1.55	&	0.33	&	1.93	&	8.86	\\
J091.4130+26.0389	&	Gaia EDR3 3430055715464804864	&	06 05 39.138	&	+26 02 20.119	&	$-$2.633	&	$-$0.218	&	+0.110	&	13.9078	&	0.0032	&	1.653195	&	2644	&	187	&	+0.186	&	0.046	&	0.56	&	+1.23	&	0.15	&	2.04	&	8.88	\\
J091.6776+11.6763	&	Gaia EDR3 3342350627854659200	&	06 06 42.643	&	+11 40 34.956	&	$-$3.318	&	$-$1.041	&	$-$0.268	&	15.4398	&	0.0062	&	2.546904	&	3488	&	439	&	+0.182	&	0.095	&	0.89	&	+1.80	&	0.27	&	1.87	&	8.73	\\
J091.7692+19.7434	&	Gaia EDR3 3374567972112361344	&	06 07 04.631	&	+19 44 36.301	&	$-$1.820	&	$-$0.334	&	$-$0.013	&	14.8044	&	0.0034	&	1.901111	&	1850	&	102	&	+0.886	&	0.224	&	1.26	&	+2.18	&	0.12	&	1.53	&	9.44	\\
J091.9664+24.5688	&	Gaia EDR3 3426334353938979712	&	06 07 51.942	&	+24 34 07.954	&	$-$1.856	&	$-$0.204	&	+0.068	&	12.7566	&	0.0032	&	2.593905	&	1868	&	80	&	+0.003	&	0.019	&	0.62	&	+0.78	&	0.09	&	2.37	&	8.65	\\
J091.9955+27.3638	&	Gaia EDR3 3430281325805346560	&	06 07 58.919	&	+27 21 49.799	&	$-$2.154	&	$-$0.144	&	+0.131	&	13.3315	&	0.0029	&	2.264348	&	2163	&	59	&	+0.198	&	0.018	&	0.54	&	+1.11	&	0.06	&	2.07	&	8.89	\\
J092.1616+30.8849	&	Gaia EDR3 3438220590031263488	&	06 08 38.791	&	+30 53 05.858	&	$-$2.628	&	$-$0.037	&	+0.244	&	14.1741	&	0.0043	&	2.167543	&	2640	&	124	&	$-$0.018	&	0.013	&	1.76	&	+0.33	&	0.10	&	2.58	&	8.63	\\
J093.1745+23.0573	&	Gaia EDR3 3425199073822806656	&	06 12 41.897	&	+23 03 26.375	&	$-$1.585	&	$-$0.226	&	+0.065	&	12.7228	&	0.0035	&	2.545983	&	1602	&	57	&	+0.035	&	0.010	&	0.62	&	+1.05	&	0.08	&	2.23	&	8.66	\\
J093.7415+41.4780	&	Gaia EDR3 960127092776565504	&	06 14 57.982	&	+41 28 40.828	&	$-$2.694	&	+0.383	&	+0.545	&	14.0878	&	0.0031	&	3.384194	&	2775	&	188	&	+0.200	&	0.012	&	0.35	&	+1.50	&	0.15	&	1.94	&	8.87	\\
J093.9930+13.8659	&	Gaia EDR3 3344570095153338880	&	06 15 58.324	&	+13 51 57.366	&	$-$1.992	&	$-$0.593	&	$-$0.050	&	13.2036	&	0.0034	&	4.428715	&	2079	&	59	&	$-$0.030	&	0.013	&	1.04	&	+0.56	&	0.06	&	2.51	&	8.60	\\
J094.2380+10.5337	&	Gaia EDR3 3329277370179300992	&	06 16 57.128	&	+10 32 01.512	&	$-$2.107	&	$-$0.751	&	$-$0.107	&	13.3359	&	0.0034	&	2.086148	&	2240	&	90	&	+0.014	&	0.011	&	0.75	&	+0.83	&	0.09	&	2.33	&	8.66	\\
J094.2801+22.2224	&	Gaia EDR3 3377000229273262464	&	06 17 07.246	&	+22 13 20.828	&	$-$2.498	&	$-$0.411	&	+0.126	&	15.2448	&	0.0037	&	2.216659	&	2535	&	238	&	+0.270	&	0.040	&	1.69	&	+1.49	&	0.20	&	1.92	&	8.95	\\
J094.3323+15.5243	&	Gaia EDR3 3345042270976679296	&	06 17 19.752	&	+15 31 27.517	&	$-$2.875	&	$-$0.785	&	$-$0.015	&	14.5985	&	0.0034	&	2.773137	&	2980	&	229	&	+0.186	&	0.091	&	1.56	&	+0.63	&	0.17	&	2.26	&	8.85	\\
J094.7507+19.5193	&	Gaia EDR3 3373985986866233344	&	06 19 00.180	&	+19 31 09.652	&	$-$2.794	&	$-$0.591	&	+0.097	&	13.7843	&	0.0035	&	3.477698	&	2858	&	503	&	+0.190	&	0.179	&	0.98	&	+0.46	&	0.38	&	2.34	&	8.83	\\
J094.9673+47.4169	&	Gaia EDR3 968928992593888896	&	06 19 52.158	&	+47 25 01.153	&	$-$1.201	&	+0.282	&	+0.323	&	13.7027	&	0.0031	&	0.942541	&	1276	&	29	&	+0.449	&	0.007	&	0.27	&	+2.89	&	0.05	&	1.51	&	8.35	\\
J095.0108+01.8715	&	Gaia EDR3 3124515178002448640	&	06 20 02.603	&	+01 52 17.560	&	$-$2.081	&	$-$1.091	&	$-$0.252	&	13.7858	&	0.0035	&	2.379236	&	2363	&	91	&	+0.114	&	0.015	&	1.10	&	+0.81	&	0.08	&	2.23	&	8.81	\\
J095.1623+09.4416	&	Gaia EDR3 3328904635738842240	&	06 20 38.975	&	+09 26 29.908	&	$-$3.151	&	$-$1.211	&	$-$0.145	&	13.0599	&	0.0031	&	2.680946	&	3379	&	203	&	+0.021	&	0.038	&	0.62	&	$-$0.20	&	0.13	&	2.77	&	8.64	\\
J095.4959+23.6564	&	Gaia EDR3 3377580290376100864	&	06 21 59.020	&	+23 39 23.154	&	$-$1.878	&	$-$0.284	&	+0.150	&	13.7218	&	0.0040	&	2.412959	&	1905	&	68	&	+0.076	&	0.034	&	1.41	&	+0.90	&	0.08	&	2.23	&	8.76	\\
J096.8934+14.9451	&	Gaia EDR3 3356762034138431488	&	06 27 34.421	&	+14 56 42.663	&	$-$4.422	&	$-$1.347	&	+0.131	&	14.9600	&	0.0035	&	1.445579	&	4624	&	548	&	$-$0.015	&	0.171	&	0.98	&	+0.64	&	0.26	&	2.45	&	8.62	\\
J097.1763+16.4813	&	Gaia EDR3 3369363914860764032	&	06 28 42.318	&	+16 28 52.975	&	$-$2.907	&	$-$0.818	&	+0.136	&	13.9936	&	0.0031	&	3.245951	&	3023	&	187	&	+0.080	&	0.041	&	0.64	&	+0.94	&	0.13	&	2.21	&	8.76	\\
J097.4242+24.5808	&	Gaia EDR3 3383547992815165568	&	06 29 41.808	&	+24 34 50.984	&	$-$2.180	&	$-$0.329	&	+0.251	&	12.6711	&	0.0031	&	3.152217	&	2219	&	110	&	+0.130	&	0.015	&	0.53	&	+0.42	&	0.11	&	2.38	&	8.80	\\
J097.6873+03.0070	&	Gaia EDR3 3124278752939435136	&	06 30 44.966	&	+03 00 25.394	&	$-$1.967	&	$-$1.041	&	$-$0.125	&	13.3766	&	0.0036	&	1.633668	&	2229	&	261	&	+0.087	&	0.143	&	1.14	&	+0.47	&	0.25	&	2.39	&	8.77	\\
J097.8891-02.7981	&	Gaia EDR3 3117203739213080320	&	06 31 33.393	&	-02 47 53.445	&	$-$1.408	&	$-$0.920	&	$-$0.168	&	13.4002	&	0.0055	&	2.500117	&	1690	&	53	&	+0.111	&	0.028	&	0.88	&	+1.34	&	0.07	&	2.05	&	8.76	\\
J097.9021-12.7568	&	Gaia EDR3 2952057092722928512	&	06 31 36.519	&	-12 45 24.652	&	$-$3.947	&	$-$3.575	&	$-$0.950	&	15.1616	&	0.0031	&	4.403193	&	5410	&	668	&	+0.096	&	0.008	&	0.62	&	+0.84	&	0.26	&	2.23	&	8.79	\\
J098.3181-07.1966	&	Gaia EDR3 3006903756370947200	&	06 33 16.365	&	-07 11 48.013	&	$-$1.032	&	$-$0.786	&	$-$0.166	&	12.5661	&	0.0033	&	1.777259	&	1308	&	20	&	+0.017	&	0.008	&	1.14	&	+0.84	&	0.03	&	2.32	&	8.67	\\
J098.5294-00.6734	&	Gaia EDR3 3119219178384615680	&	06 34 07.071	&	-00 40 24.297	&	$-$1.124	&	$-$0.691	&	$-$0.096	&	12.2876	&	0.0031	&	1.756423	&	1323	&	23	&	+0.009	&	0.033	&	0.89	&	+0.71	&	0.04	&	2.38	&	8.67	\\
J098.5904+17.0987	&	Gaia EDR3 3370947211604288512	&	06 34 21.697	&	+17 05 55.655	&	$-$3.429	&	$-$0.969	&	+0.253	&	14.6223	&	0.0032	&	1.494752	&	3572	&	271	&	+0.068	&	0.080	&	0.80	&	+1.05	&	0.16	&	2.18	&	8.74	\\
J098.6389+19.1051	&	Gaia EDR3 3371482669469578752	&	06 34 33.350	&	+19 06 18.596	&	$-$2.284	&	$-$0.570	&	+0.206	&	12.9173	&	0.0031	&	3.274034	&	2363	&	90	&	+0.042	&	0.010	&	0.56	&	+0.48	&	0.08	&	2.43	&	8.72	\\
J098.6603-02.9952	&	Gaia EDR3 3105147388056955648	&	06 34 38.494	&	-02 59 42.730	&	$-$3.244	&	$-$2.163	&	$-$0.348	&	15.4711	&	0.0035	&	3.230228	&	3914	&	527	&	+0.308	&	0.199	&	1.21	&	+1.27	&	0.30	&	1.97	&	8.98	\\
J098.8005+06.4102	&	Gaia EDR3 3132150087966065280	&	06 35 12.130	&	+06 24 36.783	&	$-$2.084	&	$-$0.989	&	$-$0.027	&	13.6585	&	0.0041	&	5.645326	&	2307	&	74	&	+0.220	&	0.310	&	0.90	&	+0.94	&	0.07	&	2.13	&	8.90	\\
J098.8719+18.1926	&	Gaia EDR3 3371169106793702912	&	06 35 29.270	&	+18 11 33.380	&	$-$3.106	&	$-$0.828	&	+0.270	&	13.6022	&	0.0035	&	2.237143	&	3225	&	172	&	$-$0.001	&	0.070	&	0.57	&	+0.50	&	0.12	&	2.48	&	8.66	\\
J099.0745+18.8221	&	Gaia EDR3 3371413537674000896	&	06 36 17.897	&	+18 49 19.635	&	$-$1.714	&	$-$0.442	&	+0.163	&	13.0924	&	0.0031	&	2.237610	&	1777	&	45	&	+0.105	&	0.011	&	0.44	&	+1.40	&	0.05	&	2.04	&	8.73	\\
J099.2222+00.8030	&	Gaia EDR3 3120576662929339520	&	06 36 53.343	&	+00 48 10.826	&	$-$1.730	&	$-$1.021	&	$-$0.101	&	13.4288	&	0.0037	&	8.348570	&	2011	&	79	&	$-$0.080	&	0.133	&	1.76	&	+0.15	&	0.09	&	2.82	&	8.49	\\
J099.6300+18.6987	&	Gaia EDR3 3371580839539015168	&	06 38 31.210	&	+18 41 55.351	&	$-$3.316	&	$-$0.876	&	+0.340	&	14.9543	&	0.0035	&	0.729200	&	3446	&	422	&	+0.069	&	0.028	&	0.81	&	+1.42	&	0.26	&	2.07	&	8.62	\\
J099.6853+17.3396	&	Gaia EDR3 3359006648466147968	&	06 38 44.489	&	+17 20 22.562	&	$-$2.112	&	$-$0.607	&	+0.196	&	12.9434	&	0.0030	&	1.722808	&	2206	&	162	&	+0.041	&	0.036	&	0.67	&	+0.54	&	0.16	&	2.40	&	8.72	\\
J100.1756+00.2310	&	Gaia EDR3 3119551127818713984	&	06 40 42.143	&	+00 13 51.695	&	$-$1.811	&	$-$1.110	&	$-$0.085	&	13.7892	&	0.0034	&	1.461506	&	2126	&	92	&	+0.082	&	0.071	&	1.05	&	+1.09	&	0.09	&	2.16	&	8.75	\\
J100.3063+00.9026	&	Gaia EDR3 3125802538617129344	&	06 41 13.520	&	+00 54 09.510	&	$-$2.780	&	$-$1.668	&	$-$0.105	&	14.2538	&	0.0032	&	2.400151	&	3244	&	223	&	$-$0.113	&	0.080	&	1.90	&	$-$0.21	&	0.15	&	3.12	&	8.40	\\
J100.6219+02.4457	&	Gaia EDR3 3126824087999901184	&	06 42 29.256	&	+02 26 44.784	&	$-$2.745	&	$-$1.568	&	$-$0.048	&	15.2004	&	0.0041	&	2.672425	&	3162	&	320	&	+0.185	&	0.159	&	1.64	&	+1.05	&	0.22	&	2.10	&	8.88	\\
J100.7339+03.0338	&	Gaia EDR3 3127282000233416704	&	06 42 56.155	&	+03 02 01.854	&	$-$1.765	&	$-$0.989	&	$-$0.018	&	13.8472	&	0.0037	&	1.305920	&	2023	&	82	&	+0.057	&	0.113	&	1.29	&	+1.02	&	0.09	&	2.21	&	8.72	\\
J101.0527+00.4465	&	Gaia EDR3 3125571400658095872	&	06 44 12.650	&	+00 26 47.729	&	$-$3.649	&	$-$2.254	&	$-$0.105	&	14.9604	&	0.0034	&	2.284570	&	4290	&	444	&	$-$0.027	&	0.261	&	1.21	&	+0.54	&	0.22	&	2.51	&	8.61	\\
J101.1640-09.7055	&	Gaia EDR3 2954359023396877952	&	06 44 39.365	&	-09 42 20.120	&	$-$4.209	&	$-$3.637	&	$-$0.575	&	15.2700	&	0.0038	&	4.103059	&	5592	&	913	&	+0.106	&	0.334	&	0.61	&	+0.91	&	0.35	&	2.20	&	8.80	\\
J101.1743+21.6307	&	Gaia EDR3 3378235530585937280	&	06 44 41.849	&	+21 37 50.549	&	$-$3.952	&	$-$0.898	&	+0.589	&	13.9466	&	0.0032	&	2.928838	&	4095	&	446	&	+0.006	&	0.010	&	0.28	&	+0.58	&	0.23	&	2.44	&	8.67	\\
J101.3758+01.7599	&	Gaia EDR3 3125978082521883520	&	06 45 30.194	&	+01 45 35.835	&	$-$1.676	&	$-$0.995	&	$-$0.018	&	13.1192	&	0.0033	&	1.278825	&	1949	&	65	&	+0.001	&	0.009	&	0.73	&	+0.93	&	0.07	&	2.32	&	8.61	\\
J101.4384+09.9330	&	Gaia EDR3 3350693726383641984	&	06 45 45.230	&	+09 55 59.142	&	$-$2.943	&	$-$1.276	&	+0.183	&	14.4764	&	0.0033	&	5.071231	&	3213	&	240	&	+0.251	&	0.035	&	0.77	&	+1.14	&	0.16	&	2.04	&	8.93	\\
J101.5831+07.1368	&	Gaia EDR3 3133176860025981568	&	06 46 19.948	&	+07 08 12.660	&	$-$2.637	&	$-$1.286	&	+0.109	&	14.3717	&	0.0029	&	1.899468	&	2936	&	240	&	+0.098	&	0.033	&	0.84	&	+1.16	&	0.18	&	2.12	&	8.77	\\
J102.5257-00.7905	&	Gaia EDR3 3113138501131391104	&	06 50 06.184	&	-00 47 26.063	&	$-$3.769	&	$-$2.493	&	$-$0.052	&	15.1167	&	0.0032	&	1.532218	&	4519	&	657	&	+0.207	&	0.168	&	1.30	&	+0.51	&	0.31	&	2.31	&	8.85	\\
J102.8631-08.4771	&	Gaia EDR3 3050995856277736192	&	06 51 27.165	&	-08 28 37.532	&	$-$1.784	&	$-$1.523	&	$-$0.158	&	14.1116	&	0.0028	&	3.637115	&	2351	&	107	&	+0.215	&	0.010	&	0.82	&	+1.43	&	0.10	&	1.96	&	8.90	\\
J102.9289-19.4359	&	Gaia EDR3 2933202014498055936	&	06 51 42.948	&	-19 26 09.558	&	$-$0.803	&	$-$0.970	&	$-$0.192	&	13.9002	&	0.0029	&	0.770109	&	1273	&	26	&	+0.474	&	0.012	&	0.54	&	+2.81	&	0.04	&	1.49	&	8.93	\\
\hline
\end{tabular}                                                                                                                                                                   
\end{adjustbox}
\end{table*}
\setcounter{table}{0}
\begin{table*}
\caption{Essential data for our sample stars, sorted by increasing right ascension. The columns denote: (1) ATLAS object identifier. (2) Gaia EDR3 identifier. (3) Right ascension (J2000; Gaia EDR3). (4) Declination (J2000; Gaia EDR3). (5) X coordinate towards the Galactic centre. (6) Y coordinate in direction of Galactic rotation. (7) Z coordinate towards the north Galactic pole. (8) Magnitude in the $G$ band (Gaia EDR3). (9) $G$\,mag error. (10) Rotational period (ATLAS). (11) Median of the photogeometric distance posterior,  $D$ (Gaia EDR3). (12) Distance error. (13) Dereddened colour index ($BP-RP$)${_0}$ (Gaia EDR3). (14) Colour index error. (15) Absorption in the $G$ band, $A_G$. (16) Intrinsic absolute magnitude in the $G$ band, M${_G}{_0}$. (17) Absolute magnitude error. (18) Mass. (19) Logarithmic age.}
\label{table_master_part3}
\begin{adjustbox}{max width=\textwidth,angle=90}
\begin{tabular}{lllllllllllllllllll}
\hline
\hline
(1) & (2) & (3) & (4) & (5) & (6) & (7) & (8) & (9) & (10) & (11) & (12) & (13) & (14) & (15) & (16) & (17) & (18) & (19) \\
\hline
ID\_ATO	&	ID\_EDR3	&	RA(J2000)	&	Dec(J2000)	&	X	&	Y	&	Z	&	$G$	&	e\_$G$	&	$P$\_rot	&	$D$	&	e\_$D$	&	($BP-RP$)${_0}$	&	e\_($BP-RP$)${_0}$	&	$A_G$	&	M${_G}{_0}$	&	e\_M${_G}{_0}$	&	Mass	&	$\log t$ \\
	&		&	(hh mm ss.sss)	&	(dd mm ss.sss)	&	(kpc)	&	(kpc)	&	(kpc)	&	(mag)	&	(mag)	&	(d)	&	(pc)	&	(pc)	&	(mag)	&	(mag)	&	(mag)	&	(mag)	&	(mag)	&	(Msun)	&	(yrs) \\
\hline
\hline
J103.3079+10.5510	&	Gaia EDR3 3159021155714729472	&	06 53 13.898	&	+10 33 03.368	&	$-$2.613	&	$-$1.148	&	+0.259	&	13.5394	&	0.0035	&	1.542512	&	2865	&	181	&	+0.029	&	0.044	&	0.20	&	+1.06	&	0.14	&	2.23	&	8.65	\\
J103.8302-21.8643	&	Gaia EDR3 2925856280391227392	&	06 55 19.270	&	-21 51 51.860	&	$-$1.340	&	$-$1.776	&	$-$0.351	&	14.7461	&	0.0030	&	1.685637	&	2252	&	91	&	+0.427	&	0.028	&	0.33	&	+2.63	&	0.09	&	1.55	&	8.89	\\
J103.8576-00.4954	&	Gaia EDR3 3112497245331079936	&	06 55 25.828	&	-00 29 43.490	&	$-$3.195	&	$-$2.141	&	+0.044	&	13.7489	&	0.0031	&	1.958371	&	3846	&	284	&	+0.030	&	0.298	&	0.79	&	+0.03	&	0.16	&	2.64	&	8.67	\\
J104.5104-01.8353	&	Gaia EDR3 3112035828406024960	&	06 58 02.496	&	-01 50 07.370	&	$-$2.557	&	$-$1.812	&	+0.035	&	14.9317	&	0.0033	&	2.534619	&	3134	&	309	&	+0.111	&	0.283	&	0.69	&	+1.74	&	0.21	&	1.94	&	8.53	\\
J105.0327+06.1169	&	Gaia EDR3 3129646018951520512	&	07 00 07.860	&	+06 07 01.002	&	$-$3.113	&	$-$1.688	&	+0.292	&	13.9587	&	0.0034	&	3.901667	&	3553	&	236	&	+0.124	&	0.019	&	0.62	&	+0.56	&	0.14	&	2.32	&	8.80	\\
J105.3860-21.3214	&	Gaia EDR3 2928971334276898944	&	07 01 32.642	&	-21 19 17.200	&	$-$0.803	&	$-$1.071	&	$-$0.174	&	13.6448	&	0.0028	&	0.800163	&	1350	&	27	&	+0.537	&	0.009	&	0.41	&	+2.58	&	0.04	&	1.50	&	9.24	\\
J105.5347-01.9077	&	Gaia EDR3 3109062371007803776	&	07 02 08.333	&	-01 54 27.955	&	$-$2.662	&	$-$1.923	&	+0.087	&	13.8002	&	0.0036	&	2.944283	&	3285	&	149	&	$-$0.029	&	0.048	&	0.64	&	+0.57	&	0.10	&	2.51	&	8.60	\\
J105.6391-13.1320	&	Gaia EDR3 3044811343886708608	&	07 02 33.393	&	-13 07 55.376	&	$-$2.646	&	$-$2.729	&	$-$0.234	&	15.1346	&	0.0029	&	1.537205	&	3808	&	410	&	+0.118	&	0.044	&	0.96	&	+1.27	&	0.24	&	2.07	&	8.79	\\
J105.9212-15.5797	&	Gaia EDR3 2936164889096440192	&	07 03 41.100	&	-15 34 47.208	&	$-$0.884	&	$-$0.988	&	$-$0.102	&	12.5845	&	0.0030	&	1.616409	&	1329	&	77	&	+0.192	&	0.011	&	0.51	&	+1.44	&	0.13	&	1.97	&	8.87	\\
J105.9294+02.6272	&	Gaia EDR3 3115630406794434176	&	07 03 43.056	&	+02 37 38.235	&	$-$2.351	&	$-$1.468	&	+0.190	&	14.1671	&	0.0031	&	2.226779	&	2778	&	204	&	+0.066	&	0.027	&	0.39	&	+1.54	&	0.16	&	2.04	&	8.52	\\
J106.2127-00.9740	&	Gaia EDR3 3109566875045011584	&	07 04 51.051	&	-00 58 26.651	&	$-$1.803	&	$-$1.278	&	+0.098	&	13.6232	&	0.0035	&	2.027484	&	2212	&	73	&	+0.072	&	0.025	&	0.34	&	+1.52	&	0.07	&	2.04	&	8.56	\\
J106.4471-03.7806	&	Gaia EDR3 3107804838943634304	&	07 05 47.322	&	-03 46 50.281	&	$-$2.865	&	$-$2.233	&	+0.093	&	14.3678	&	0.0034	&	2.670424	&	3634	&	319	&	+0.013	&	0.047	&	0.58	&	+0.97	&	0.19	&	2.28	&	8.63	\\
J106.5580+18.8987	&	Gaia EDR3 3361893764140100096	&	07 06 13.942	&	+18 53 55.622	&	$-$6.284	&	$-$1.983	&	+1.359	&	14.9067	&	0.0037	&	3.271069	&	6728	&	979	&	$-$0.077	&	0.012	&	0.07	&	+0.67	&	0.31	&	2.61	&	8.40	\\
J106.7195-14.1516	&	Gaia EDR3 3044457576020575744	&	07 06 52.687	&	-14 09 05.914	&	$-$1.738	&	$-$1.882	&	$-$0.137	&	14.8783	&	0.0032	&	2.774142	&	2566	&	434	&	+0.280	&	0.111	&	1.10	&	+1.69	&	0.37	&	1.85	&	8.95	\\
J107.2480-12.7673	&	Gaia EDR3 3044897483752096256	&	07 08 59.543	&	-12 46 02.633	&	$-$1.900	&	$-$1.986	&	$-$0.094	&	14.1566	&	0.0033	&	2.951689	&	2750	&	142	&	+0.276	&	0.029	&	1.16	&	+0.80	&	0.11	&	2.16	&	8.92	\\
J108.1121-08.8579	&	Gaia EDR3 3048321122442630400	&	07 12 26.921	&	-08 51 28.542	&	$-$1.354	&	$-$1.272	&	+0.019	&	13.1451	&	0.0030	&	1.944000	&	1858	&	46	&	+0.157	&	0.006	&	0.43	&	+1.37	&	0.05	&	2.01	&	8.83	\\
J108.1707-08.2617	&	Gaia EDR3 3051756477804650624	&	07 12 40.978	&	-08 15 42.222	&	$-$2.223	&	$-$2.051	&	+0.048	&	14.0242	&	0.0029	&	2.079303	&	3025	&	188	&	$-$0.014	&	0.133	&	1.11	&	+0.50	&	0.13	&	2.50	&	8.63	\\
J108.2723-11.0753	&	Gaia EDR3 3045550765453488768	&	07 13 05.355	&	-11 04 31.191	&	$-$2.121	&	$-$2.139	&	$-$0.016	&	15.4240	&	0.0037	&	1.732496	&	3012	&	257	&	+0.056	&	0.162	&	2.71	&	+0.27	&	0.18	&	2.50	&	8.72	\\
J108.3588-12.4968	&	Gaia EDR3 3045104535537115008	&	07 13 26.123	&	-12 29 48.781	&	$-$1.965	&	$-$2.074	&	$-$0.044	&	14.4735	&	0.0029	&	3.078432	&	2857	&	122	&	+0.116	&	0.014	&	0.85	&	+1.34	&	0.09	&	2.05	&	8.77	\\
J108.6795-07.3062	&	Gaia EDR3 3051966621964045952	&	07 14 43.083	&	-07 18 22.381	&	$-$2.096	&	$-$1.893	&	+0.089	&	14.3276	&	0.0030	&	1.828566	&	2825	&	142	&	+0.087	&	0.074	&	0.56	&	+1.49	&	0.11	&	2.03	&	8.64	\\
J108.7050-20.4362	&	Gaia EDR3 2930001812880632576	&	07 14 49.204	&	-20 26 10.416	&	$-$1.052	&	$-$1.434	&	$-$0.133	&	13.2960	&	0.0028	&	2.359520	&	1784	&	38	&	+0.212	&	0.008	&	0.53	&	+1.51	&	0.05	&	1.93	&	8.89	\\
J108.7797-22.5861	&	Gaia EDR3 2927954590949675904	&	07 15 07.147	&	-22 35 10.314	&	$-$2.126	&	$-$3.116	&	$-$0.342	&	14.2677	&	0.0032	&	1.460651	&	3787	&	319	&	+0.084	&	0.024	&	0.74	&	+0.63	&	0.18	&	2.33	&	8.77	\\
J109.4314-15.7080	&	Gaia EDR3 3031253300366764416	&	07 17 43.550	&	-15 42 28.802	&	$-$2.265	&	$-$2.687	&	$-$0.089	&	14.9541	&	0.0036	&	3.587642	&	3516	&	344	&	+0.195	&	0.250	&	1.41	&	+0.82	&	0.21	&	2.18	&	8.87	\\
J109.5054-14.0571	&	Gaia EDR3 3032637546843653376	&	07 18 01.315	&	-14 03 25.575	&	$-$1.902	&	$-$2.146	&	$-$0.031	&	12.6786	&	0.0029	&	1.948277	&	2868	&	152	&	$-$0.036	&	0.011	&	0.59	&	$-$0.20	&	0.12	&	2.87	&	8.57	\\
J109.7734-07.1470	&	Gaia EDR3 3054942037866252288	&	07 19 05.632	&	-07 08 49.261	&	$-$2.478	&	$-$2.267	&	+0.167	&	13.9320	&	0.0031	&	1.858906	&	3363	&	264	&	$-$0.039	&	0.050	&	0.44	&	+0.84	&	0.17	&	2.44	&	8.50	\\
J110.1723-23.4818	&	Gaia EDR3 5617891650669025792	&	07 20 41.375	&	-23 28 54.704	&	$-$1.227	&	$-$1.895	&	$-$0.176	&	14.9461	&	0.0029	&	1.657834	&	2265	&	114	&	+0.318	&	0.509	&	1.73	&	+1.43	&	0.11	&	1.92	&	9.00	\\
J110.2057-08.5343	&	Gaia EDR3 3048588028890791936	&	07 20 49.371	&	-08 32 03.657	&	$-$2.187	&	$-$2.103	&	+0.136	&	15.0582	&	0.0044	&	2.501694	&	3038	&	190	&	+0.270	&	0.130	&	0.36	&	+2.26	&	0.14	&	1.71	&	8.61	\\
J110.2507-12.0207	&	Gaia EDR3 3034664500832358912	&	07 21 00.190	&	-12 01 14.838	&	$-$2.664	&	$-$2.855	&	+0.066	&	14.9108	&	0.0036	&	1.607134	&	3906	&	365	&	+0.024	&	0.074	&	0.98	&	+0.96	&	0.20	&	2.27	&	8.66	\\
J110.2675-03.2520	&	Gaia EDR3 3059901071523913856	&	07 21 04.208	&	-03 15 07.462	&	$-$1.565	&	$-$1.278	&	+0.180	&	12.5848	&	0.0057	&	2.809982	&	2029	&	92	&	+0.114	&	0.026	&	0.01	&	+0.99	&	0.10	&	2.16	&	8.81	\\
J110.3902-17.5039	&	Gaia EDR3 3027827978047324544	&	07 21 33.657	&	-17 30 14.402	&	$-$1.792	&	$-$2.285	&	$-$0.075	&	14.8237	&	0.0035	&	2.618199	&	2905	&	208	&	+0.033	&	0.361	&	1.49	&	+1.00	&	0.16	&	2.25	&	8.67	\\
J110.4158-24.8742	&	Gaia EDR3 5616860377480409344	&	07 21 39.816	&	-24 52 27.459	&	$-$1.476	&	$-$2.401	&	$-$0.242	&	14.7029	&	0.0031	&	1.275878	&	2829	&	169	&	+0.062	&	0.063	&	1.06	&	+1.37	&	0.13	&	2.10	&	8.62	\\
J110.4771-08.9354	&	Gaia EDR3 3047764735202190464	&	07 21 54.524	&	-08 56 07.674	&	$-$2.339	&	$-$2.287	&	+0.150	&	14.4830	&	0.0031	&	2.377864	&	3275	&	183	&	+0.197	&	0.031	&	0.51	&	+1.38	&	0.12	&	1.98	&	8.88	\\
J110.9074-12.0800	&	Gaia EDR3 3034627426673874304	&	07 23 37.800	&	-12 04 48.109	&	$-$1.741	&	$-$1.889	&	+0.068	&	13.5939	&	0.0031	&	1.647087	&	2569	&	115	&	+0.099	&	0.072	&	0.71	&	+0.83	&	0.10	&	2.24	&	8.79	\\
J110.9088+00.6468	&	Gaia EDR3 3110806948062742656	&	07 23 38.127	&	+00 38 48.538	&	$-$0.768	&	$-$0.559	&	+0.124	&	12.0645	&	0.0045	&	6.048553	&	958	&	21	&	+0.202	&	0.029	&	0.76	&	+1.38	&	0.05	&	1.98	&	8.89	\\
J110.9392-21.9535	&	Gaia EDR3 5619699316501817984	&	07 23 45.411	&	-21 57 12.817	&	$-$1.500	&	$-$2.228	&	$-$0.146	&	14.2400	&	0.0032	&	1.609184	&	2690	&	121	&	+0.053	&	0.034	&	1.32	&	+0.75	&	0.10	&	2.31	&	8.74	\\
J111.1535-20.9269	&	Gaia EDR3 5619923926116738432	&	07 24 36.844	&	-20 55 37.074	&	$-$1.747	&	$-$2.518	&	$-$0.132	&	13.3500	&	0.0029	&	1.573938	&	3067	&	103	&	+0.064	&	0.050	&	0.53	&	+0.38	&	0.07	&	2.45	&	8.74	\\
J111.2165-23.1396	&	Gaia EDR3 5619369669171589632	&	07 24 51.977	&	-23 08 22.642	&	$-$2.174	&	$-$3.376	&	$-$0.242	&	14.6298	&	0.0031	&	7.770775	&	4023	&	305	&	+0.051	&	0.029	&	1.65	&	$-$0.05	&	0.16	&	2.66	&	8.68	\\
J112.1996-20.1412	&	Gaia EDR3 5620435508258520832	&	07 28 47.925	&	-20 08 28.481	&	$-$1.388	&	$-$1.984	&	$-$0.052	&	13.3988	&	0.0030	&	5.587132	&	2422	&	60	&	+0.050	&	0.005	&	0.65	&	+0.82	&	0.05	&	2.28	&	8.73	\\
J112.7672-19.4071	&	Gaia EDR3 5716567630938745088	&	07 31 04.144	&	-19 24 25.681	&	$-$1.429	&	$-$2.013	&	$-$0.017	&	13.3513	&	0.0031	&	1.769319	&	2468	&	70	&	+0.072	&	0.078	&	0.75	&	+0.62	&	0.06	&	2.34	&	8.76	\\
J112.8463-20.1548	&	Gaia EDR3 5620395135564902400	&	07 31 23.120	&	-20 09 17.513	&	$-$1.247	&	$-$1.802	&	$-$0.027	&	13.5636	&	0.0029	&	1.345138	&	2192	&	119	&	+0.095	&	0.008	&	0.67	&	+1.16	&	0.12	&	2.12	&	8.77	\\
J113.2387-19.4340	&	Gaia EDR3 5716581031237970304	&	07 32 57.300	&	-19 26 02.616	&	$-$1.468	&	$-$2.086	&	$-$0.001	&	13.3008	&	0.0030	&	2.273453	&	2551	&	74	&	$-$0.002	&	0.043	&	1.17	&	+0.09	&	0.06	&	2.66	&	8.64	\\
J113.5306-15.0738	&	Gaia EDR3 3029887810007115264	&	07 34 07.361	&	-15 04 25.861	&	$-$2.373	&	$-$2.950	&	+0.154	&	14.5531	&	0.0031	&	2.449479	&	3789	&	268	&	+0.166	&	0.076	&	0.85	&	+0.80	&	0.15	&	2.20	&	8.85	\\
J113.7045-18.5902	&	Gaia EDR3 5716740529142761472	&	07 34 49.091	&	-18 35 24.839	&	$-$1.268	&	$-$1.767	&	+0.029	&	13.6235	&	0.0029	&	3.116609	&	2175	&	67	&	+0.433	&	0.175	&	0.63	&	+1.29	&	0.07	&	1.94	&	9.05	\\
J113.8870-24.3497	&	Gaia EDR3 5615293573412352256	&	07 35 32.884	&	-24 20 59.355	&	$-$1.732	&	$-$2.936	&	$-$0.112	&	14.6048	&	0.0033	&	1.424931	&	3411	&	251	&	+0.054	&	0.043	&	1.05	&	+0.88	&	0.16	&	2.26	&	8.73	\\
J114.2059-18.3864	&	Gaia EDR3 5717113641542052864	&	07 36 49.431	&	-18 23 11.234	&	$-$1.869	&	$-$2.610	&	+0.072	&	15.3662	&	0.0030	&	1.379164	&	3211	&	290	&	+0.315	&	0.179	&	1.45	&	+1.37	&	0.20	&	1.94	&	8.99	\\
J114.4350-24.3432	&	Gaia EDR3 5615128474874397440	&	07 37 44.412	&	-24 20 35.581	&	$-$1.324	&	$-$2.265	&	$-$0.066	&	14.8710	&	0.0032	&	2.242821	&	2625	&	184	&	+0.215	&	0.026	&	1.64	&	+1.11	&	0.15	&	2.07	&	8.90	\\
J114.4848-21.4294	&	Gaia EDR3 5619270747487432448	&	07 37 56.376	&	-21 25 46.118	&	$-$1.625	&	$-$2.520	&	+0.001	&	13.8852	&	0.0035	&	2.972624	&	2998	&	132	&	+0.083	&	0.179	&	0.75	&	+0.74	&	0.10	&	2.28	&	8.77	\\
J114.8681-18.8011	&	Gaia EDR3 5716848899758250624	&	07 39 28.367	&	-18 48 04.202	&	$-$1.770	&	$-$2.534	&	+0.088	&	14.3776	&	0.0040	&	2.485193	&	3092	&	170	&	+0.121	&	0.211	&	1.56	&	+0.36	&	0.12	&	2.41	&	8.78	\\
J114.9662-29.1402	&	Gaia EDR3 5599763590149196032	&	07 39 51.901	&	-29 08 24.806	&	$-$0.905	&	$-$1.865	&	$-$0.123	&	13.0419	&	0.0029	&	1.842183	&	2077	&	60	&	+0.114	&	0.026	&	0.43	&	+1.02	&	0.06	&	2.15	&	8.81	\\
J115.0928-10.6198	&	Gaia EDR3 3040276064580646912	&	07 40 22.289	&	-10 37 11.301	&	$-$2.127	&	$-$2.364	&	+0.324	&	14.3393	&	0.0033	&	1.908329	&	3196	&	217	&	+0.021	&	0.012	&	0.06	&	+1.77	&	0.15	&		&		\\
J115.2178-25.5994	&	Gaia EDR3 5614180592769077248	&	07 40 52.286	&	-25 35 58.010	&	$-$1.490	&	$-$2.704	&	$-$0.078	&	13.4778	&	0.0030	&	2.097984	&	3088	&	114	&	+0.013	&	0.012	&	0.66	&	+0.36	&	0.08	&	2.52	&	8.68	\\
J115.2306-17.0845	&	Gaia EDR3 5717317085555220096	&	07 40 55.354	&	-17 05 04.405	&	$-$2.021	&	$-$2.755	&	+0.165	&	14.2894	&	0.0030	&	1.214488	&	3420	&	196	&	+0.034	&	0.057	&	0.69	&	+0.92	&	0.12	&	2.27	&	8.69	\\
J115.2403-24.8545	&	Gaia EDR3 5614295526096577920	&	07 40 57.689	&	-24 51 16.299	&	$-$1.858	&	$-$3.284	&	$-$0.070	&	15.1178	&	0.0030	&	1.893434	&	3774	&	275	&	+0.201	&	0.029	&	0.83	&	+1.38	&	0.16	&	1.98	&	8.89	\\
J116.0922-28.4220	&	Gaia EDR3 5599913639125075328	&	07 44 22.133	&	-28 25 19.543	&	$-$1.557	&	$-$3.190	&	$-$0.135	&	15.0275	&	0.0032	&	2.682821	&	3553	&	326	&	+0.183	&	0.089	&	1.10	&	+1.14	&	0.20	&	2.07	&	8.87	\\
J116.2532-21.3016	&	Gaia EDR3 5712207758099475840	&	07 45 00.771	&	-21 18 05.881	&	$-$1.211	&	$-$1.929	&	+0.060	&	13.6462	&	0.0029	&	3.280127	&	2278	&	104	&	+0.156	&	0.155	&	0.45	&	+1.40	&	0.10	&	2.00	&	8.82	\\
J116.3042-11.2203	&	Gaia EDR3 3037077310375003648	&	07 45 13.016	&	-11 13 13.089	&	$-$2.376	&	$-$2.746	&	+0.418	&	13.7169	&	0.0032	&	2.061657	&	3655	&	329	&	$-$0.063	&	0.064	&	0.33	&	+0.57	&	0.20	&	2.60	&	8.49	\\
J116.3339-28.7092	&	Gaia EDR3 5599847015590653184	&	07 45 20.140	&	-28 42 33.382	&	$-$1.899	&	$-$3.951	&	$-$0.163	&	15.5904	&	0.0033	&	1.514279	&	4387	&	509	&	+0.167	&	0.044	&	1.32	&	+1.04	&	0.25	&	2.11	&	8.86	\\
J116.4925-16.0440	&	Gaia EDR3 5718996005449601536	&	07 45 58.219	&	-16 02 38.495	&	$-$2.117	&	$-$2.854	&	+0.269	&	14.5629	&	0.0031	&	1.807299	&	3564	&	302	&	+0.161	&	0.045	&	0.07	&	+1.72	&	0.18	&	1.91	&	8.72	\\
J116.5342-19.8451	&	Gaia EDR3 5715755744681588608	&	07 46 08.216	&	-19 50 42.441	&		&		&		&	13.9075	&	0.0059	&	1.860305	&		&	0	&		&		&		&		&		&		&		\\
J116.5376-20.4285	&	Gaia EDR3 5715650638239585920	&	07 46 09.041	&	-20 25 42.645	&	$-$1.408	&	$-$2.190	&	+0.099	&	13.9076	&	0.0029	&	1.682906	&	2606	&	105	&	+0.057	&	0.124	&	0.66	&	+1.16	&	0.09	&	2.16	&	8.69	\\
J116.6244-18.2192	&	Gaia EDR3 5716944213671349760	&	07 46 29.880	&	-18 13 09.334	&	$-$1.949	&	$-$2.824	&	+0.201	&	14.4120	&	0.0032	&	1.615868	&	3438	&	239	&	+0.189	&	0.037	&	0.53	&	+1.19	&	0.15	&	2.05	&	8.88	\\
J116.6529-30.1024	&	Gaia EDR3 5598677203938894080	&	07 46 36.699	&	-30 06 08.772	&	$-$1.605	&	$-$3.553	&	$-$0.176	&	14.3816	&	0.0038	&	2.273473	&	3902	&	201	&	+0.102	&	0.034	&	0.63	&	+0.77	&	0.11	&	2.26	&	8.79	\\
J116.8586-29.7166	&	Gaia EDR3 5598969845827526528	&	07 47 26.081	&	-29 42 59.926	&	$-$1.246	&	$-$2.727	&	$-$0.117	&	14.2228	&	0.0033	&	2.245591	&	3000	&	136	&	+0.268	&	0.018	&	1.24	&	+0.57	&	0.10	&	2.26	&	8.88	\\
J117.1588-19.3073	&	Gaia EDR3 5715872224194437632	&	07 48 38.137	&	-19 18 26.576	&	$-$1.791	&	$-$2.715	&	+0.185	&	14.1542	&	0.0034	&	2.776477	&	3258	&	209	&	+0.209	&	0.011	&	0.38	&	+1.22	&	0.14	&	2.03	&	8.90	\\
J117.4249-26.8279	&	Gaia EDR3 5601626472083501696	&	07 49 41.982	&	-26 49 40.491	&	$-$1.701	&	$-$3.368	&	$-$0.023	&	14.5315	&	0.0040	&	2.784376	&	3773	&	270	&	+0.192	&	0.116	&	0.57	&	+1.07	&	0.15	&	2.09	&	8.88	\\
J117.6125-26.0800	&	Gaia EDR3 5602249379782658944	&	07 50 27.013	&	-26 04 48.074	&	$-$1.323	&	$-$2.557	&	+0.009	&	14.1958	&	0.0030	&	4.750825	&	2880	&	165	&	+0.169	&	0.016	&	0.48	&	+1.40	&	0.12	&	1.99	&	8.84	\\
J117.7143-28.9873	&	Gaia EDR3 5600519225204862336	&	07 50 51.455	&	-28 59 14.317	&	$-$1.909	&	$-$4.128	&	$-$0.097	&	14.3923	&	0.0029	&	1.505937	&	4549	&	337	&	+0.099	&	0.015	&	1.06	&	+0.03	&	0.16	&	2.58	&	8.73	\\
J117.8237-09.5101	&	Gaia EDR3 3040695974941812736	&	07 51 17.702	&	-09 30 36.468	&	$-$1.246	&	$-$1.402	&	+0.287	&	12.9492	&	0.0033	&	1.789376	&	1898	&	125	&	$-$0.069	&	0.012	&	0.15	&	+1.42	&	0.14	&		&		\\
J118.1364-26.0263	&	Gaia EDR3 5602054422627227520	&	07 52 32.749	&	-26 01 35.059	&	$-$2.099	&	$-$4.091	&	+0.049	&	14.1629	&	0.0030	&	3.159168	&	4598	&	275	&	+0.051	&	0.065	&	0.86	&	$-$0.02	&	0.13	&	2.64	&	8.69	\\
J118.1717-28.2232	&	Gaia EDR3 5600675394523048448	&	07 52 41.220	&	-28 13 23.699	&	$-$1.425	&	$-$3.020	&	$-$0.029	&	14.6811	&	0.0030	&	1.836496	&	3339	&	172	&	+0.049	&	0.011	&	1.13	&	+0.92	&	0.11	&	2.25	&	8.72	\\
\hline
\end{tabular}                                                                                                                                                                   
\end{adjustbox}
\end{table*}
\setcounter{table}{0}
\begin{table*}
\caption{Essential data for our sample stars, sorted by increasing right ascension. The columns denote: (1) ATLAS object identifier. (2) Gaia EDR3 identifier. (3) Right ascension (J2000; Gaia EDR3). (4) Declination (J2000; Gaia EDR3). (5) X coordinate towards the Galactic centre. (6) Y coordinate in direction of Galactic rotation. (7) Z coordinate towards the north Galactic pole. (8) Magnitude in the $G$ band (Gaia EDR3). (9) $G$\,mag error. (10) Rotational period (ATLAS). (11) Median of the photogeometric distance posterior,  $D$ (Gaia EDR3). (12) Distance error. (13) Dereddened colour index ($BP-RP$)${_0}$ (Gaia EDR3). (14) Colour index error. (15) Absorption in the $G$ band, $A_G$. (16) Intrinsic absolute magnitude in the $G$ band, M${_G}{_0}$. (17) Absolute magnitude error. (18) Mass. (19) Logarithmic age.}
\label{table_master_part4}
\begin{adjustbox}{max width=\textwidth,angle=90}
\begin{tabular}{lllllllllllllllllll}
\hline
\hline
(1) & (2) & (3) & (4) & (5) & (6) & (7) & (8) & (9) & (10) & (11) & (12) & (13) & (14) & (15) & (16) & (17) & (18) & (19) \\
\hline
ID\_ATO	&	ID\_EDR3	&	RA(J2000)	&	Dec(J2000)	&	X	&	Y	&	Z	&	$G$	&	e\_$G$	&	$P$\_rot	&	$D$	&	e\_$D$	&	($BP-RP$)${_0}$	&	e\_($BP-RP$)${_0}$	&	$A_G$	&	M${_G}{_0}$	&	e\_M${_G}{_0}$	&	Mass	&	$\log t$ \\
	&		&	(hh mm ss.sss)	&	(dd mm ss.sss)	&	(kpc)	&	(kpc)	&	(kpc)	&	(mag)	&	(mag)	&	(d)	&	(pc)	&	(pc)	&	(mag)	&	(mag)	&	(mag)	&	(mag)	&	(mag)	&	(Msun)	&	(yrs) \\
\hline
\hline
J118.3027-29.5272	&	Gaia EDR3 5600243389532206464	&	07 53 12.664	&	-29 31 38.226	&	$-$1.475	&	$-$3.300	&	$-$0.067	&	15.2374	&	0.0031	&	3.258970	&	3616	&	307	&	+0.160	&	0.090	&	0.82	&	+1.60	&	0.18	&	1.94	&	8.78	\\
J118.6780-30.9643	&	Gaia EDR3 5595430895857226240	&	07 54 42.730	&	-30 57 51.688	&	$-$0.944	&	$-$2.258	&	$-$0.065	&	13.7257	&	0.0034	&	1.914314	&	2448	&	103	&	+0.076	&	0.011	&	0.79	&	+0.97	&	0.09	&	2.20	&	8.76	\\
J118.9841-27.5769	&	Gaia EDR3 5601086233914919424	&	07 55 56.190	&	-27 34 37.034	&	$-$1.197	&	$-$2.515	&	+0.022	&	13.7855	&	0.0031	&	1.201602	&	2785	&	111	&	+0.147	&	0.008	&	0.57	&	+0.98	&	0.09	&	2.15	&	8.84	\\
J119.0168-28.8107	&	Gaia EDR3 5600407835242829568	&	07 56 04.039	&	-28 48 38.798	&	$-$1.922	&	$-$4.242	&	$-$0.013	&	14.9111	&	0.0034	&	1.298072	&	4657	&	461	&	$-$0.017	&	0.075	&	0.86	&	+0.69	&	0.21	&	2.44	&	8.61	\\
J120.3173-30.6071	&	Gaia EDR3 5596989900267016704	&	08 01 16.155	&	-30 36 25.783	&	$-$1.231	&	$-$3.009	&	$-$0.008	&	14.3856	&	0.0031	&	1.635983	&	3251	&	155	&	+0.277	&	0.017	&	0.56	&	+1.25	&	0.10	&	1.99	&	8.96	\\
J120.9589-17.2768	&	Gaia EDR3 5715047315593195008	&	08 03 50.154	&	-17 16 36.772	&	$-$1.100	&	$-$1.675	&	+0.260	&	14.2869	&	0.0030	&	0.769542	&	2021	&	225	&	+0.577	&	0.005	&	0.01	&	+2.73	&	0.24	&	1.45	&	9.30	\\
J121.7666-30.2091	&	Gaia EDR3 5596596103307222144	&	08 07 03.994	&	-30 12 32.899	&	$-$1.224	&	$-$3.040	&	+0.065	&	15.3611	&	0.0039	&	1.623933	&	3278	&	233	&	+0.189	&	0.077	&	1.07	&	+1.69	&	0.15	&	1.90	&	8.80	\\
J121.7955-28.6294	&	Gaia EDR3 5597610093546202752	&	08 07 10.930	&	-28 37 46.052	&	$-$1.185	&	$-$2.760	&	+0.105	&	13.5481	&	0.0031	&	1.831770	&	3005	&	120	&	$-$0.028	&	0.014	&	0.43	&	+0.73	&	0.09	&	2.45	&	8.57	\\
J121.8181-25.3414	&	Gaia EDR3 5697706196160845568	&	08 07 16.348	&	-25 20 29.220	&	$-$0.705	&	$-$1.445	&	+0.107	&	12.2134	&	0.0031	&	5.049751	&	1612	&	41	&	+0.050	&	0.008	&	0.06	&	+1.11	&	0.06	&	2.19	&	8.68	\\
J127.2306-24.4682	&	Gaia EDR3 5695756865121458944	&	08 28 55.360	&	-24 28 05.742	&	$-$1.307	&	$-$2.937	&	+0.472	&	13.4228	&	0.0031	&	1.760478	&	3249	&	186	&	$-$0.101	&	0.007	&	0.17	&	+0.69	&	0.12	&	2.70	&	8.23	\\
J265.2024+10.6314	&	Gaia EDR3 4492510575866960000	&	17 40 48.586	&	+10 37 53.171	&	+3.429	&	+2.371	&	+1.555	&	14.2487	&	0.0051	&	0.742114	&	4449	&	438	&	+0.552	&	0.019	&	0.32	&	+0.65	&	0.21	&	2.08	&	9.05	\\
J271.8730-19.3670	&	Gaia EDR3 4095137796471886080	&	18 07 29.530	&	-19 22 01.409	&	+1.788	&	+0.340	&	+0.016	&	13.1280	&	0.0031	&	1.379003	&	1820	&	60	&	+0.039	&	0.020	&	1.51	&	+0.30	&	0.07	&	2.51	&	8.71	\\
J275.4103-19.5093	&	Gaia EDR3 4094521726396476672	&	18 21 38.481	&	-19 30 33.576	&	+2.076	&	+0.450	&	$-$0.093	&	14.5695	&	0.0034	&	2.360844	&	2126	&	172	&	+0.021	&	0.480	&	2.78	&	+0.13	&	0.18	&	2.61	&	8.68	\\
J275.6627-12.0194	&	Gaia EDR3 4153257053415199360	&	18 22 39.051	&	-12 01 10.048	&	+1.308	&	+0.450	&	+0.020	&	12.7122	&	0.0030	&	2.767335	&	1384	&	39	&	+0.134	&	0.207	&	1.05	&	+0.89	&	0.06	&	2.19	&	8.83	\\
J276.9001-13.4054	&	Gaia EDR3 4104464919291577984	&	18 27 36.034	&	-13 24 19.463	&	+2.801	&	+0.927	&	$-$0.047	&	14.4466	&	0.0036	&	2.602447	&	2950	&	216	&	+0.344	&	0.064	&	1.98	&	+0.10	&	0.16	&	2.46	&	8.83	\\
J277.0437-19.4213	&	Gaia EDR3 4093068618757395584	&	18 28 10.509	&	-19 25 16.933	&	+1.621	&	+0.375	&	$-$0.111	&	12.8054	&	0.0048	&	1.812403	&	1667	&	51	&	+0.299	&	0.048	&	0.68	&	+1.00	&	0.07	&	2.08	&	8.95	\\
J277.6324-12.9737	&	Gaia EDR3 4104875238943201536	&	18 30 31.778	&	-12 58 25.362	&	+2.154	&	+0.743	&	$-$0.053	&	14.1326	&	0.0042	&	1.801325	&	2279	&	170	&	+0.130	&	0.184	&	1.60	&	+0.72	&	0.16	&	2.25	&	8.82	\\
J279.0944-07.2749	&	Gaia EDR3 4252839581007062272	&	18 36 22.664	&	-07 16 29.628	&	+1.625	&	+0.749	&	+0.000	&	14.2987	&	0.0038	&	3.236192	&	1789	&	72	&	+0.301	&	0.016	&	1.97	&	+1.06	&	0.09	&	2.05	&	8.96	\\
J279.6357-11.8797	&	Gaia EDR3 4106502967189674368	&	18 38 32.575	&	-11 52 47.168	&	+1.277	&	+0.488	&	$-$0.061	&	13.2288	&	0.0035	&	2.179231	&	1368	&	32	&	+0.290	&	0.011	&	1.63	&	+0.90	&	0.05	&	2.12	&	8.94	\\
J280.0484-07.0582	&	Gaia EDR3 4252761137808845696	&	18 40 11.638	&	-07 03 29.796	&	+2.288	&	+1.085	&	$-$0.032	&	13.8821	&	0.0044	&	2.216235	&	2533	&	116	&	+0.193	&	0.031	&	1.64	&	+0.26	&	0.10	&	2.43	&	8.81	\\
J282.9649-13.5589	&	Gaia EDR3 4102192778511285120	&	18 51 51.578	&	-13 33 32.154	&	+1.951	&	+0.744	&	$-$0.228	&	11.6414	&	0.0075	&	1.271750	&	2101	&	124	&	+0.317	&	0.035	&	0.71	&	$-$0.66	&	0.13	&	2.75	&	8.71	\\
J289.9475+15.3004	&	Gaia EDR3 4320708929142333568	&	19 19 47.421	&	+15 18 01.509	&	+1.770	&	+2.090	&	+0.039	&	14.2774	&	0.0036	&	2.598937	&	2739	&	220	&	+0.057	&	0.096	&	1.93	&	+0.14	&	0.17	&	2.56	&	8.71	\\
J299.4473+32.3039	&	Gaia EDR3 2034049889560166016	&	19 57 47.367	&	+32 18 14.238	&	+1.279	&	+3.298	&	+0.102	&	14.9055	&	0.0035	&	1.723220	&	3539	&	264	&	+0.032	&	0.250	&	2.19	&	$-$0.06	&	0.16	&	2.69	&	8.67	\\
J300.8633+39.6355	&	Gaia EDR3 2074018477252787584	&	20 03 27.199	&	+39 38 08.233	&	+0.698	&	+2.726	&	+0.223	&	13.2111	&	0.0031	&	2.347255	&	2823	&	155	&	$-$0.073	&	0.044	&	0.74	&	+0.20	&	0.12	&	2.78	&	8.51	\\
J301.0209+31.3865	&	Gaia EDR3 2030794854079584768	&	20 04 05.030	&	+31 23 11.788	&	+1.163	&	+2.986	&	+0.002	&	14.4842	&	0.0030	&	1.641778	&	3205	&	204	&	+0.039	&	0.149	&	1.92	&	+0.00	&	0.14	&	2.65	&	8.68	\\
J302.2996+32.2037	&	Gaia EDR3 2054874502319741824	&	20 09 11.914	&	+32 12 13.608	&	+0.558	&	+1.534	&	$-$0.012	&	13.5948	&	0.0038	&	1.178096	&	1632	&	30	&	$-$0.124	&	0.305	&	1.77	&	+0.76	&	0.04	&	2.78	&	7.87	\\
J305.4541+39.4820	&	Gaia EDR3 2061438002621031168	&	20 21 49.005	&	+39 28 55.534	&	+0.358	&	+1.609	&	+0.043	&	13.1189	&	0.0033	&	1.565190	&	1649	&	27	&	$-$0.086	&	0.011	&	2.63	&	$-$0.61	&	0.04	&	3.23	&	8.44	\\
J306.8515+38.9758	&	Gaia EDR3 2064160840081582592	&	20 27 24.383	&	+38 58 33.194	&	+0.365	&	+1.670	&	+0.009	&	14.6958	&	0.0039	&	1.381301	&	1710	&	59	&	+0.074	&	0.259	&	2.70	&	+0.82	&	0.08	&	2.26	&	8.76	\\
J306.9801+45.3863	&	Gaia EDR3 2070928849556576256	&	20 27 55.240	&	+45 23 11.037	&	+0.200	&	+1.617	&	+0.113	&	13.1856	&	0.0031	&	1.558536	&	1633	&	27	&	+0.007	&	0.008	&	1.52	&	+0.58	&	0.04	&	2.43	&	8.67	\\
J308.7048+45.8077	&	Gaia EDR3 2071178644852414080	&	20 34 49.169	&	+45 48 28.035	&	+0.241	&	+2.295	&	+0.131	&	13.5256	&	0.0034	&	0.970713	&	2311	&	65	&	$-$0.730	&	0.444	&	2.64	&	$-$0.96	&	0.06	&		&		\\
J310.1614-15.1051	&	Gaia EDR3 6887222608825741440	&	20 40 38.745	&	-15 06 18.310	&	+4.719	&	+2.828	&	$-$3.288	&	14.5233	&	0.0041	&	0.795110	&	6409	&	507	&	+0.599	&	0.015	&	0.18	&	+0.29	&	0.17	&	2.24	&	8.95	\\
J311.5037+47.0071	&	Gaia EDR3 2166724143728120832	&	20 46 00.901	&	+47 00 25.601	&	+0.137	&	+2.023	&	+0.086	&	12.6766	&	0.0035	&	2.701917	&	2030	&	40	&	$-$0.166	&	0.341	&	1.19	&	$-$0.07	&	0.04	&	3.36	&	8.14	\\
J315.2267+51.3206	&	Gaia EDR3 2169422654485439616	&	21 00 54.433	&	+51 19 14.273	&	$-$0.034	&	+1.921	&	+0.112	&	13.6476	&	0.0030	&	0.964536	&	1924	&	42	&	$-$0.231	&	0.041	&	3.74	&	$-$1.52	&	0.05	&	4.80	&	7.97	\\
J315.9822+55.3700	&	Gaia EDR3 2188897204438678400	&	21 03 55.728	&	+55 22 12.064	&	$-$0.108	&	+1.410	&	+0.140	&	14.1281	&	0.0036	&	2.644264	&	1421	&	175	&	+0.408	&	0.010	&	1.58	&	+1.73	&	0.27	&	1.79	&	9.08	\\
J317.2429+45.6406	&	Gaia EDR3 2162570394956913920	&	21 08 58.303	&	+45 38 26.272	&	+0.084	&	+2.080	&	$-$0.053	&	15.3574	&	0.0029	&	1.896461	&	2082	&	118	&	+0.597	&	0.011	&	0.95	&	+2.78	&	0.12	&	1.44	&	9.33	\\
J317.2517+45.2742	&	Gaia EDR3 2162544144116244352	&	21 09 00.431	&	+45 16 27.271	&	+0.082	&	+1.821	&	$-$0.054	&	13.0174	&	0.0032	&	1.385039	&	1824	&	43	&	+0.170	&	0.012	&	1.36	&	+0.33	&	0.05	&	2.40	&	8.81	\\
J317.6917+45.2552	&	Gaia EDR3 2162500232369128576	&	21 10 46.026	&	+45 15 18.868	&	+0.137	&	+3.295	&	$-$0.112	&	14.1725	&	0.0031	&	2.351053	&	3300	&	143	&	$-$0.084	&	0.032	&	0.87	&	+0.70	&	0.09	&	2.62	&	8.34	\\
J318.9587+44.0280	&	Gaia EDR3 1971527187299417728	&	21 15 50.111	&	+44 01 40.927	&	+0.065	&	+1.402	&	$-$0.085	&	12.1206	&	0.0029	&	1.266083	&	1406	&	21	&	$-$0.028	&	0.007	&	1.01	&	+0.36	&	0.03	&	2.59	&	8.61	\\
J319.0262+51.3139	&	Gaia EDR3 2172227852252643840	&	21 16 06.302	&	+51 18 50.250	&	$-$0.084	&	+1.851	&	+0.051	&	13.7517	&	0.0035	&	2.200647	&	1853	&	49	&	$-$0.055	&	0.051	&	2.26	&	+0.14	&	0.06	&	2.75	&	8.55	\\
J320.8569+54.0590	&	Gaia EDR3 2175923379549109760	&	21 23 25.672	&	+54 03 32.752	&	$-$0.089	&	+0.954	&	+0.046	&	12.3628	&	0.0032	&	1.274577	&	959	&	15	&	+0.345	&	0.055	&	2.13	&	+0.32	&	0.03	&	2.25	&	8.95	\\
J321.3283+53.8969	&	Gaia EDR3 2175962377854306944	&	21 25 18.802	&	+53 53 48.999	&	$-$0.158	&	+1.659	&	+0.070	&	14.0211	&	0.0033	&	1.124518	&	1668	&	39	&	$-$0.384	&	0.034	&	3.43	&	$-$0.54	&	0.05	&		&		\\
J321.3875+49.7047	&	Gaia EDR3 2171065737185099392	&	21 25 33.024	&	+49 42 17.159	&	$-$0.074	&	+1.662	&	$-$0.018	&	14.5533	&	0.0039	&	2.509597	&	1664	&	51	&	+0.207	&	0.015	&	1.77	&	+1.65	&	0.07	&	1.90	&	8.85	\\
J321.4270+50.1748	&	Gaia EDR3 2171181392057223552	&	21 25 42.506	&	+50 10 29.340	&	$-$0.100	&	+1.977	&	$-$0.010	&	15.4800	&	0.0033	&	2.424447	&	1979	&	100	&	+0.551	&	0.064	&	2.50	&	+1.49	&	0.11	&	1.85	&	9.14	\\
J322.5803+48.1366	&	Gaia EDR3 1978543239722001280	&	21 30 19.285	&	+48 08 12.140	&	$-$0.081	&	+2.305	&	$-$0.092	&	13.3729	&	0.0034	&	3.722594	&	2308	&	80	&	+0.044	&	0.008	&	0.65	&	+0.92	&	0.07	&	2.26	&	8.71	\\
J323.9253+55.1526	&	Gaia EDR3 2174626711751528832	&	21 35 42.077	&	+55 09 09.536	&	$-$0.222	&	+1.712	&	+0.069	&	13.6816	&	0.0031	&	2.049757	&	1728	&	36	&	$-$0.027	&	0.098	&	1.50	&	+0.98	&	0.05	&	2.36	&	8.49	\\
J326.5224+55.9721	&	Gaia EDR3 2175008998199747328	&	21 46 05.397	&	+55 58 19.737	&	$-$0.466	&	+2.933	&	+0.101	&	15.2394	&	0.0046	&	1.824793	&	2971	&	233	&	+0.318	&	0.092	&	2.19	&	+0.68	&	0.17	&	2.20	&	8.92	\\
J330.5591+53.1647	&	Gaia EDR3 1981280890545613312	&	22 02 14.211	&	+53 09 52.980	&	$-$0.310	&	+1.930	&	$-$0.058	&	14.5700	&	0.0039	&	1.767390	&	1956	&	65	&	+0.434	&	0.037	&	1.29	&	+1.81	&	0.07	&	1.76	&	9.10	\\
J331.2729+57.0606	&	Gaia EDR3 2198299575056422400	&	22 05 05.515	&	+57 03 38.338	&	$-$0.591	&	+2.836	&	+0.060	&	14.0264	&	0.0034	&	4.639860	&	2898	&	138	&	+0.024	&	0.116	&	2.00	&	$-$0.31	&	0.10	&	2.83	&	8.62	\\
J336.4884+56.9915	&	Gaia EDR3 2007771905224191744	&	22 25 57.231	&	+56 59 29.517	&	$-$0.967	&	+3.853	&	$-$0.032	&	14.8466	&	0.0034	&	2.408172	&	3973	&	269	&	$-$0.040	&	0.123	&	1.72	&	+0.12	&	0.15	&	2.72	&	8.58	\\
J336.7031+54.9666	&	Gaia EDR3 2005131187536539648	&	22 26 48.770	&	+54 58 00.005	&	$-$0.618	&	+2.654	&	$-$0.107	&	13.7046	&	0.0035	&	3.014172	&	2727	&	100	&	+0.070	&	0.019	&	0.89	&	+0.62	&	0.08	&	2.34	&	8.76	\\
J337.4678+55.8598	&	Gaia EDR3 2006717164338438784	&	22 29 52.283	&	+55 51 35.552	&	$-$0.997	&	+4.010	&	$-$0.123	&	15.4165	&	0.0040	&	1.727613	&	4134	&	472	&	+0.010	&	0.077	&	1.70	&	+0.63	&	0.25	&	2.41	&	8.67	\\
J340.1084+58.6023	&	Gaia EDR3 2008171921301411456	&	22 40 26.026	&	+58 36 08.349	&	$-$0.552	&	+1.856	&	$-$0.001	&	13.9309	&	0.0029	&	2.775725	&	1936	&	51	&	+0.190	&	0.033	&	1.33	&	+1.14	&	0.06	&	2.07	&	8.88	\\
J344.9302+56.6191	&	Gaia EDR3 2009813285998755072	&	22 59 43.260	&	+56 37 08.991	&	$-$0.931	&	+2.869	&	$-$0.157	&	14.3272	&	0.0033	&	0.955319	&	3020	&	143	&	+0.050	&	0.114	&	1.11	&	+0.80	&	0.10	&	2.29	&	8.73	\\
J350.3833+56.4514	&	Gaia EDR3 1997279574991800576	&	23 21 32.008	&	+56 27 05.270	&	$-$0.791	&	+2.092	&	$-$0.167	&	13.3329	&	0.0033	&	2.346321	&	2243	&	61	&	+0.024	&	0.015	&	1.08	&	+0.49	&	0.06	&	2.45	&	8.70	\\
\hline
\end{tabular}
\end{adjustbox}
\end{table*}

\section{Distribution of our sample stars in the Galactic [XYZ] plane}\label{appendix_xyz_plot}

\begin{figure*}
        \includegraphics[width=2.0\columnwidth,bb=0 0 600 450]{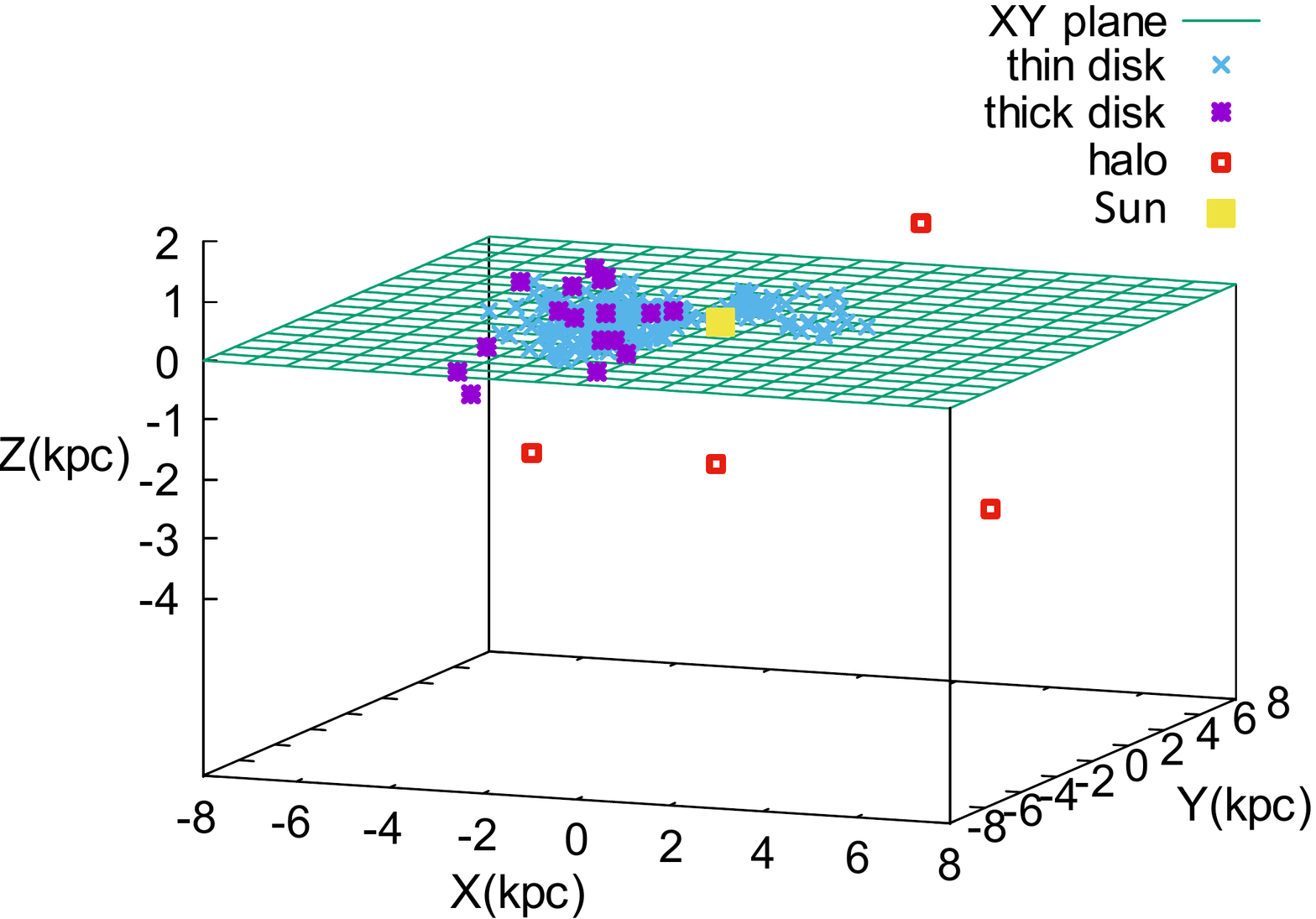}
    \caption{Distribution of our sample stars in the Galactic [XYZ] plane. Stars were assigned as probable members of the thin disk, the thick disk and the halo according to the scale heights provided by \citet{2001MNRAS.322..426O} and \citet{2017MNRAS.470.2113A}.}
		\label{map_3D}
\end{figure*}

\section{Light curves}\label{light_curves}

This section provides a sample page with the light curves of the first 16 stars of our sample, folded with the periods listed in Table \ref{table_master_part1}. The full set of light curves is available online.

\begin{figure*}
    \includegraphics[width=0.33\textwidth]{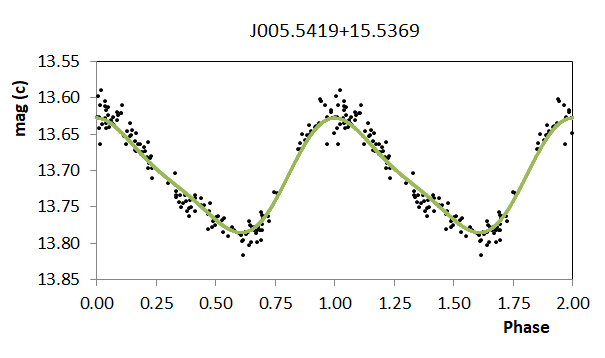}
    \includegraphics[width=0.33\textwidth]{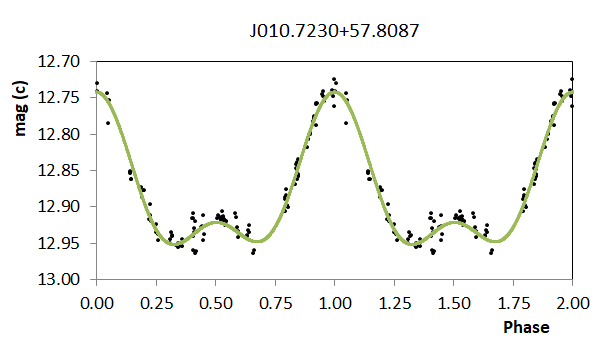}
    \includegraphics[width=0.33\textwidth]{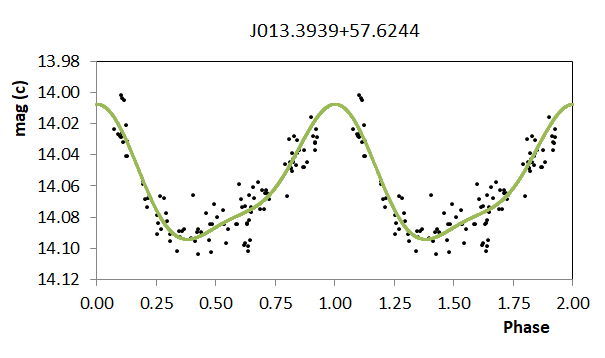}
    \includegraphics[width=0.33\textwidth]{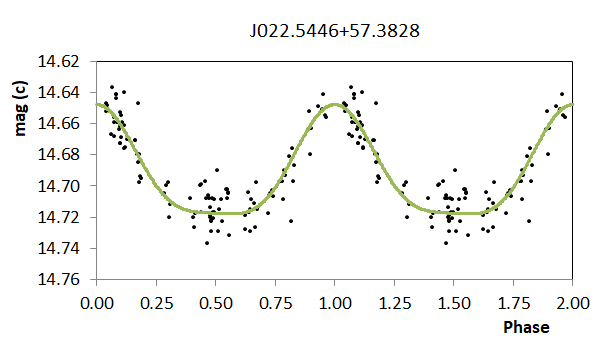}
    \includegraphics[width=0.33\textwidth]{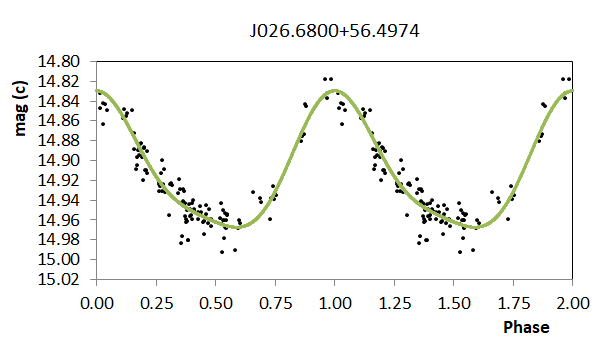}
    \includegraphics[width=0.33\textwidth]{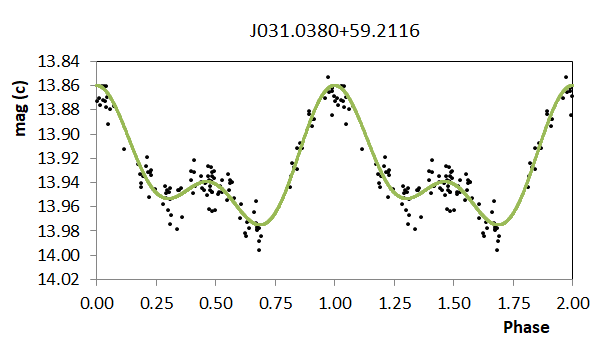}
    \includegraphics[width=0.33\textwidth]{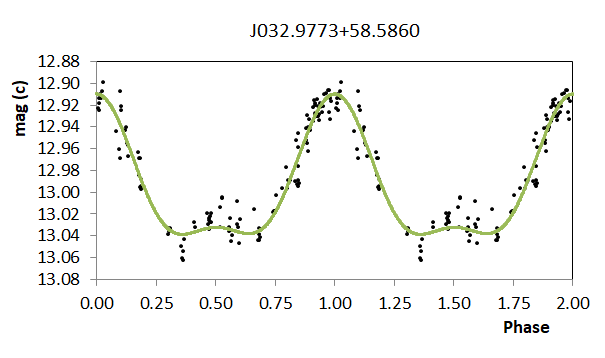}
    \includegraphics[width=0.33\textwidth]{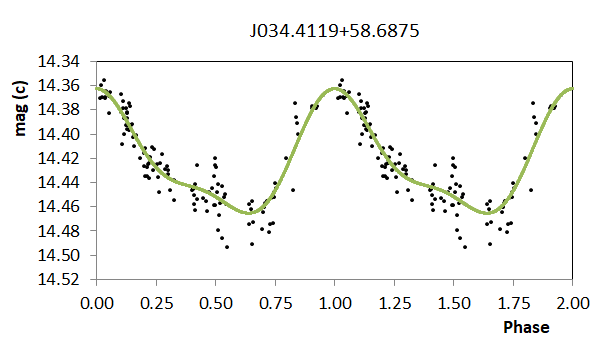}
    \includegraphics[width=0.33\textwidth]{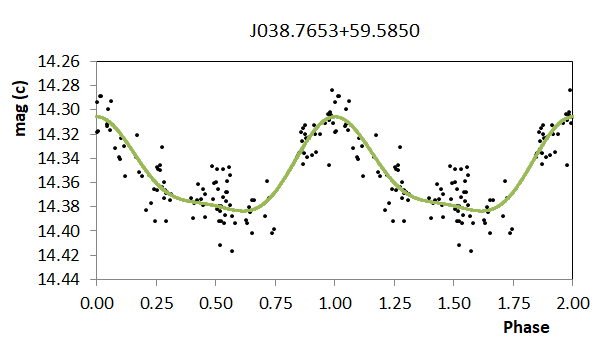}
    \includegraphics[width=0.33\textwidth]{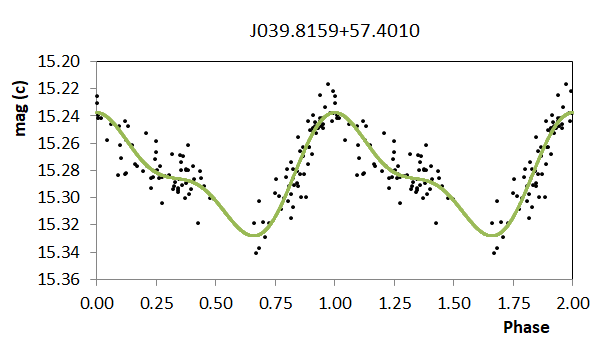}
    \includegraphics[width=0.33\textwidth]{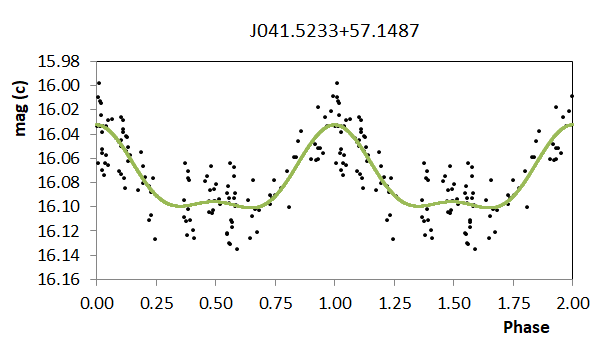}
    \includegraphics[width=0.33\textwidth]{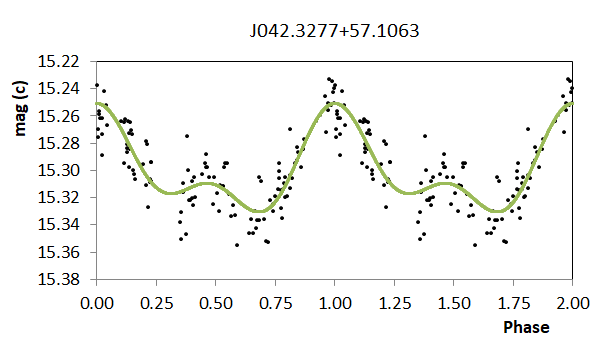}
    \includegraphics[width=0.33\textwidth]{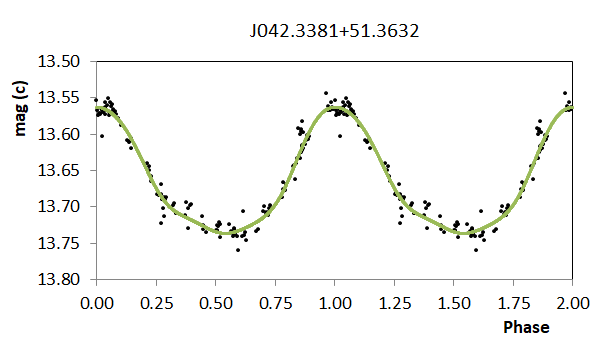}
    \includegraphics[width=0.33\textwidth]{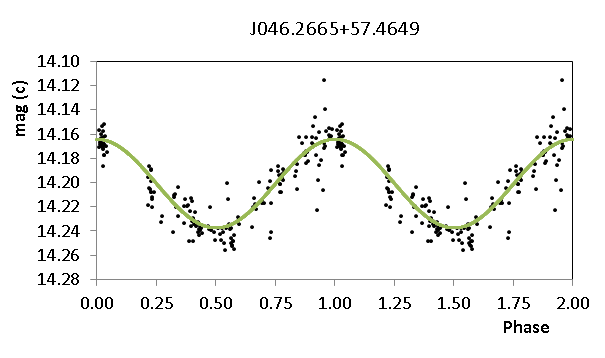}
    \includegraphics[width=0.33\textwidth]{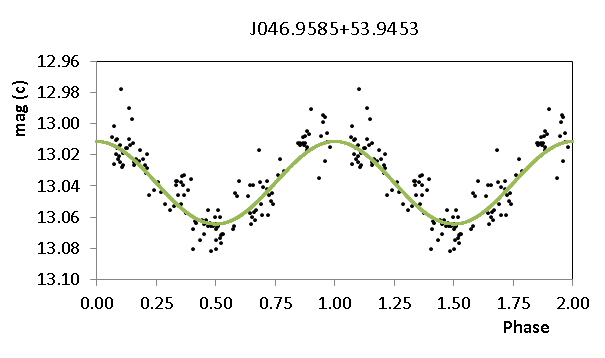}
    \includegraphics[width=0.33\textwidth]{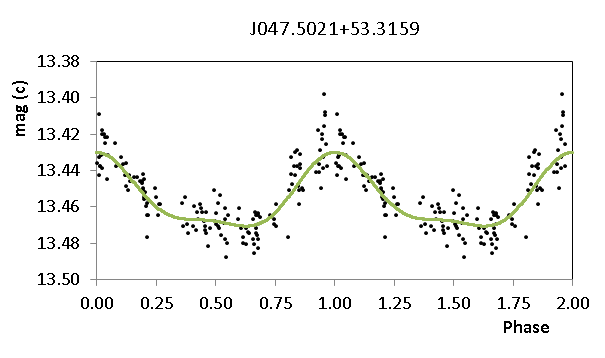}
    \includegraphics[width=0.33\textwidth]{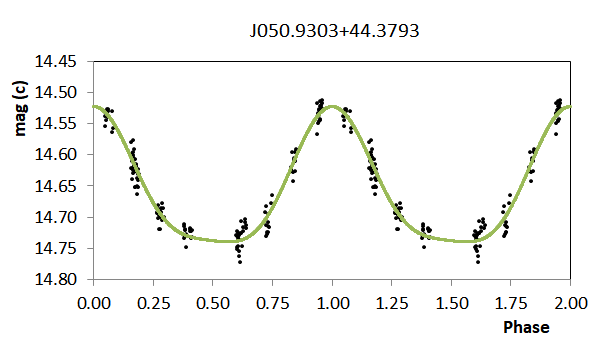}
    \includegraphics[width=0.33\textwidth]{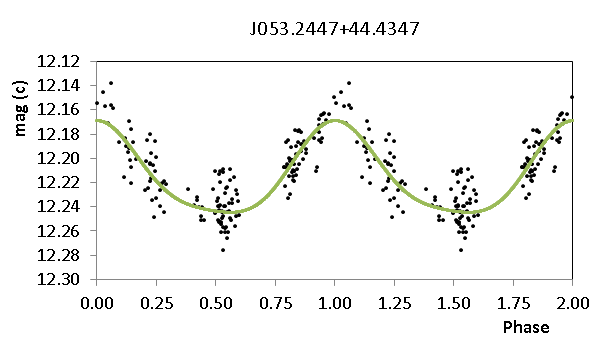}
    \caption{Sample page presenting the light curves of the first 16 objects in Table \ref{table_master_part1}, folded with the periods listed therein. The full set of light curves is available online.}
		\label{LC_sample_page}
\end{figure*}


\bsp	
\label{lastpage}
\end{document}